\RequirePackage[l2tabu, orthodox]{nag}
\RequirePackage{fix-cm}
\documentclass[11pt,onecolumn,letter]{IEEEtran}
 \usepackage{etoolbox}
\patchcmd{\section}{\centering}{}{}{}

\usepackage{geometry}
\usepackage{graphicx} 

\usepackage[cmex10]{amsmath} 
\usepackage{amssymb}
\usepackage{amsfonts}
\usepackage{mathrsfs} 

\usepackage{cuted}

\usepackage{subcaption}

\usepackage[noadjust]{cite}
\usepackage{epstopdf}

\usepackage[x11names]{xcolor}   

\usepackage[disable,colorinlistoftodos]{todonotes}  

\usepackage[T1]{fontenc}
\usepackage{charter}
\usepackage{environ}
\usepackage{tikz}
\usetikzlibrary{calc,matrix}

\usepackage[section]{placeins} 
\usepackage{stackengine}

\usepackage{enumitem}
\setlist[enumerate]{label=\roman*} 

\usepackage{microtype}

\newtheorem{theorem}{Theorem}

\usepackage[colorlinks]{hyperref}
\hypersetup{%
	colorlinks = true,
	citecolor=Green4,
	linkcolor  = blue
}

\pagebreak

\ifCLASSINFOpdf
\else
\fi

\begin{document}
\IEEEoverridecommandlockouts

\title{Sensing Method for Two-Target Detection in Time-Constrained Vector Gaussian Channel}


	\author{\IEEEauthorblockN{Muhammad Fahad\IEEEauthorrefmark{1},~\IEEEmembership{Member,~IEEE}, and Daniel R. Fuhrmann\IEEEauthorrefmark{2},~\IEEEmembership{Fellow,~IEEE}} \\
		\IEEEauthorblockA{Department of Applied Computing,\\
			Michigan Technological University\\ 
			Houghton, MI 49931, USA \\
			\IEEEauthorrefmark{1}mfahad@mtu.edu,
			\IEEEauthorrefmark{2}fuhrmann@mtu.edu}}
\maketitle
\begin{abstract} This paper considers a vector Gaussian channel of fixed identity covariance matrix and binary input signalling as the mean of it. A linear transformation is performed on the vector input signal. The objective is to find the optimal scaling matrix, under the total time constraint, that would: i) maximize the mutual information between the input and output random vectors, ii) maximize the MAP detection. It was found that the two metrics lead to different optimal solutions for our experimental design problem. We have used the Monte Carlo method for our computational work.
\end{abstract}

%
%
  
\section{Introduction}  \label{intro}
In \cite{guo2005mutual} a Gaussian channel $Y|X \sim \mathcal{N}(\sqrt{T} \cdot X,1) $ is considered and discovered that mutual information $I(T)$ is concave function in $T$ for arbitrary input distribution. Whereas in \cite{atar2012mutual} the Poisson channel $Y|X \sim \operatorname{Poiss}(T \cdot X) $ is investigated and found a result similar to the Gaussian channel: $I(T)$ is concave in $T$ for arbitrary input distribution. In \cite {paper1}\cite{phdfahad} it was observed that concavity of $I$ looked preserved under linear time constraint in a vector Poisson channel with a $2-$long binary input signalling and a $3-$long conditionally Poisson vector. It was further observed from a computational viewpoint that MAP detector was not necessarily reaching to the same optimal argument as that was given by mutual information. Here we construct an analogous model to that of Poisson channel such that at least the concavity of $I$ remains intact for the Gaussian channel too. Compared to vector Poisson channel, the literature on vector Gaussian channel is comparatively richer and may help in providing some insight into the Poisson channel.

In past work \cite{wang2014bregman} a generalization of Bregman divergence is developed to unify the vector Gaussian and Poisson channel models from the perspective of the gradient of mutual information; and mutual information is considered for signal recovery and classification with an energy constraint $\mathtt{Tr}(\Phi \Phi^{\intercal})=1$. MAP estimation is used for the classification purpose in \cite{wang2014bregman} using Monte Carlo method to first approximate the gradient and then gradient descent is employed for the classification problem. It was noted in \cite{wang2014bregman}  that mutual information well served the classification problem. In this paper we attempted to use the detection theory criterion for signal classification and then compared with the information theoretic solution.
Another work \cite{palomar2006gradient} provides some results relevant to Gaussian channels about the concavity of $I$ w.r.t squared singular values of the scaling matrix when certain conditions on the channel covariance and precoder matrix are satisfied. It is found that for our problem $I$ is concave in affine space defined by $(T_1,T_2,T_3):= $ \allowbreak $(\frac{T-T_3}{2},\frac{T-T_3}{2},T_3) $ where $ 0 \le T_3 \le T $. 
For a Gaussian channel $Y|X \sim \mathcal{N}(\Phi X,\sigma^2_n I)$ with input $X \sim \mathcal{N}(0, \Sigma)$ where $\Sigma$ is full rank covariance matrix; then the two solutions from maximizing the mutual information and minimizing the mean-square error in scaling matrix under the power constraint $\mathtt{Tr}(\Phi \Phi^{\intercal})$ leads to the same optimal solution which is a water-filling power allocation i-e concentrate more power resource to modes that provide higher snr \cite{yang2007mimo}. Our problem is different in the input signalling, and we took the detection theory criterion instead of the estimation theory (MMSE) and then compared the optimal solution with one obtained from $I$ using Monte Carlo computational method. 
\begin{figure}[t]
		\centering
		\includegraphics[width=0.7\linewidth]{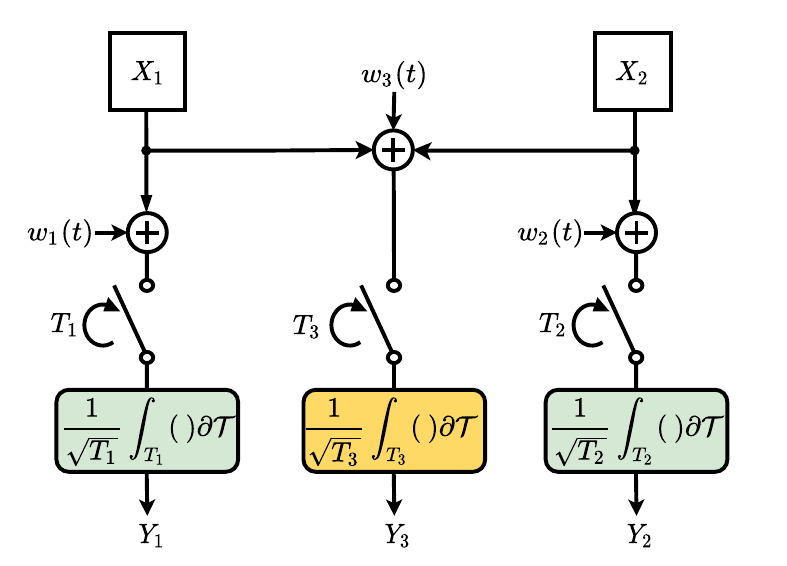}
		\caption{Illustration of sensing paradigm for detection of $2-$long hidden random vector $X$ from $3-$long observable random vector $Y$ through a vector Gaussian channel under a total time constraint  $T=\sum_{i=1}^{3} T_i$. $w_i(t)$ are independent white Gaussian noise processes. Only one of the integrators becomes active for a time $T_i$ such that time constraint is always satisfied after the total sensing time $T$ is consumed. Objective is to maximize the mutual information between input and output, $I(X_1, X_2; Y_1, Y_2, Y_3)$, and Bayes probability of total detections, $P_d$, by satisfying the time constraint. \todo[inline]{1) Open draw.io in chrome. \newline 2) Open file from C:$\backslash$Users$\backslash$mfahad$\backslash$Dropbox(4)$\backslash$Apps$\backslash$drawio$\backslash$Sensing\_Paradigm15.html \newline 3) Export to HTML and save in the same folder.  \newline 4) Open HTML and print PDF \newline 5) Crop the pdf using acrobats tools: crop  \newline 6) Save the pdf in folder C:$\backslash$Users$\backslash$mfahad$\backslash$Dropbox$\backslash$PhD\_Paper\_1 }}
		\label{Figg_6}
	\end{figure}

The rest of the paper is organized as follows: Section \ref{probdes} introduces the problem description, explaining the vector Gaussian channel under consideration. Section \ref{detofc} provides the \emph{information theoretic} model, while Section \ref{detd} describes the \emph{detection theoretic} model of the problem. Section \ref{comand} discusses the computed results from the previous two sections. Finally, Section \ref{con} concludes the paper.

 \textit{Notation:} Upper case letters denote random vectors. Realizations of the random vectors are denoted by lower case letters. A number in subscript is used to show the component number of the random vector. We use $ X_1 $ and $ Y_1 $ to represent scalar input and output random variables, respectively. The superscript $ (\cdot)^{\intercal}  $ denotes the matrix/vector transpose. $ T $ is a given finite time. $ \alpha $ is an arbitrary positive scalar variable. $ \Phi $ represents the scaling matrix. $ p $ is the prior probability and $ f_X(x) $ denotes the probability mass function of $ X $. $K-$dimensional multivariate Gaussian distribution is represented by $\mathcal{N}(x;\mu,\Sigma )=(2 \pi)^{-\frac{k}{2}} \mathtt{det}(\Sigma)^{-\frac{1}{2}} e^{-{\frac{1}{2}}(x-\mu)^\intercal \Sigma^{-1}(x-\mu)}$.
\clearpage
\section{Problem Description} \label{probdes}
  We consider the vector Gaussian channel:
	
	\begin{equation}
		\label{e0} 
		\begin{aligned}
			\begin{bmatrix}
				Y_1     \\
				Y_2  \\
				Y_3   \\
			\end{bmatrix}
			&=
			\begin{bmatrix}
				\sqrt{T_1} & 0  \\
				0 & \sqrt{T_2}  \\
				\sqrt{T_3} & \sqrt{T_3} \\
			\end{bmatrix} 
			\begin{bmatrix}
				X_1   \\
				X_2   \\
			\end{bmatrix} +
			\begin{bmatrix}
				N_1  \\
				N_2  \\
				N_3 \\
			\end{bmatrix},\\
		\end{aligned}
	\end{equation}	
with $X_1$ and $X_2$ be two independent and identical distributed (i.i.d) transformed Bernoulli random variables with $ p $ being the probability of occurance of $ 1 $. We consider probability mass function $ f $ of discrete random vector $ X \equiv [X_1,X_2]^{\intercal}$ as
\begin{IEEEeqnarray*}{rCl}
	f_{X}(x)=\left\{
	\begin{array}{ll}
		p^2 & \quad x  =  \: [\lambda_1 \quad \lambda_1]^{\intercal}  \\ 
		(1-p)^2 & \quad x  =  \: [\lambda_0 \quad \lambda_0]^{\intercal} \\
		p(1-p) & \quad  x  =  \: [\lambda_0\quad \lambda_1]^{\intercal} \quad \text{or} \quad  x  =  \: [\lambda_1 \quad \lambda_0]^{\intercal}  \\             
	\end{array}
	\right. \label{w8} \IEEEyesnumber
\end{IEEEeqnarray*}
Noise vector $N$ is a multivariate Gaussian with zero mean and identity covariance matrix; and independent of input $X$. The constraint on the scaling matrix is $T_1+T_2+T_3=T$. 
The conditional distribution of vector $Y$ given $X$ is a multivariate Gaussian:
\begin{equation}
	\label{e01} 
	\begin{aligned}
		Y\Bigg|(X=x_1 x_2)
		&\sim
		\mathcal{N} \Bigg(\begin{bmatrix}
			\sqrt{T_1} \cdot x_1   \\
			\sqrt{T_2} \cdot x_2 \\
			\sqrt{T_3} \cdot (x_1+x_2) \\
		\end{bmatrix},
		I \Bigg).\\
	\end{aligned}
\end{equation}

The objective is optimal time-allocation, $ (T_1,T_2,T_3) $, of total available time resource, $ T $, that would maximize the \textit{reward} i.e. either the mutual information or probability of total correct detections. Mathematically we may write 
\begin{IEEEeqnarray*}{rCl}
	& \underset{T_1, T_2, T_3}{\text{max}} \: I(X_1, X_2 ; Y_1, Y_2, Y_3) \: \: \text{s.t.} \: \:  T_1+T_2+T_3=T. \label{e1} \IEEEyesnumber
\end{IEEEeqnarray*} 

From the \emph{detection theoretic} aspect we maximize the Bayesian probability of total correct detections, $P_d $, of hidden random vector $ X $ from observable random vector $ Y $, as 
\begin{IEEEeqnarray*}{rCl}
	& \underset{T_1, T_2, T_3}{\text{max}} \: P_d \quad  \text{s.t.} \quad  T_1+T_2+T_3=T \label{e9} \IEEEyesnumber
\end{IEEEeqnarray*}
\section{Information Theoretic Description} \label{detofc}

 \subsection{Scalar Gaussian channel} \label{form0}   

The scalar version of the Gaussian channel is first presented, and then we extend it to the vector version. We start with mutual information between a scalar random variable $ X_1 $ which is a transformed Bernoulli random variable and $ Y_1 $ is a univariate Gaussian mixture. The probability mass function of $ Y_1 $ is then given as
\begin{IEEEeqnarray*}{rCl}
f(Y_1)	 &=& (1-p) \: \cdot \mathcal{N}(Y_1; \lambda_0 \sqrt{ T},1) + p \:  \cdot \mathcal{N}(Y_1; \lambda_1 \sqrt{ T},1).
\end{IEEEeqnarray*}
The mutual information $ I $ can be written as
\begin{IEEEeqnarray*}{rCl}
	I(X_1;Y_1) &=& H(Y_1)-H(Y_1|X_1)	
\end{IEEEeqnarray*}
 where $ H(\cdot) $ is the Shannon entropy and $ f $ is the probability mass function of random variate $ Y $ with $ \mathcal{Y} $ as the corresponding support.
 We may write differential entropy $ H(Y_1) $ as
 \begin{IEEEeqnarray*}{rCl}
 	H(Y_1)	&=&  - \int_{-\infty}^{\infty} \:  \Bigg( (1-p) \cdot  \mathcal{N}(y_1;\lambda_0 \sqrt{T}, 1) + p\cdot \mathcal{N}(y_1; \lambda_1 \sqrt{T},1)   \cdot \\ && \> \operatorname{Log}_2 \Big[(1-p) \cdot  \mathcal{N}(y_1;\lambda_0 \sqrt{T}, 1) + p\cdot \mathcal{N}(y_1; \lambda_1 \sqrt{T},1) \Big] \Bigg)   \,d{y_1}, \label{AL5} \IEEEyesnumber
 \end{IEEEeqnarray*}
 and 
 \begin{IEEEeqnarray*}{rCl}
 	H(Y_1|X_1)	&=& (1-p)\cdot 0.5 \cdot \operatorname{Log}_2[2 \pi \: e ] +p \cdot 0.5 \cdot \operatorname{Log}_2[2 \pi \: e ]. \label{AL6} \IEEEyesnumber
 \end{IEEEeqnarray*}

In the following section we formulate the mutual information expression for our vector Gaussian model.
   \begin{figure}[t] 
	\centering
	\includegraphics[width=0.6\linewidth]{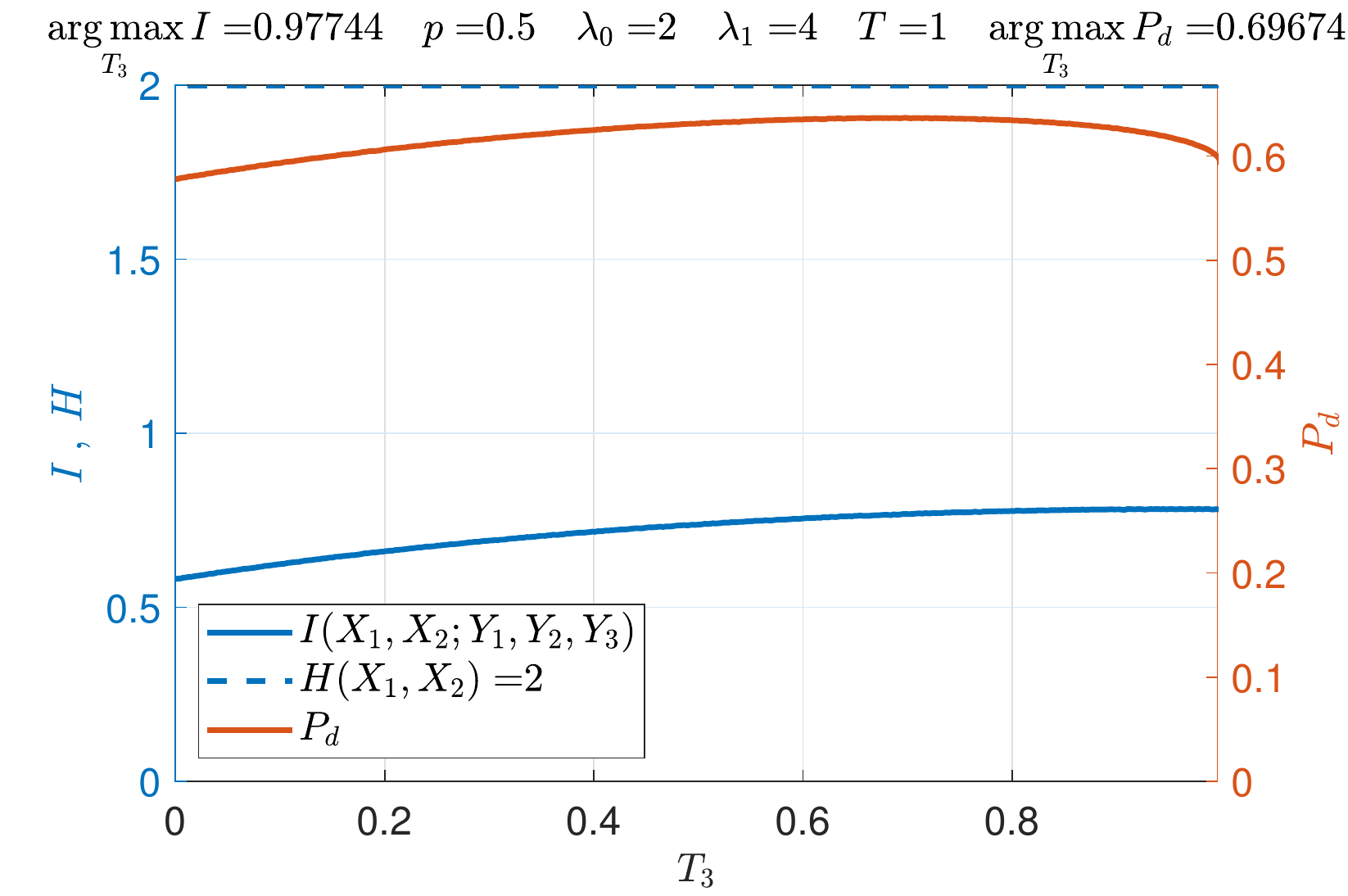}  
	\caption{Mutual information $ I(X;Y) $ vs. $T_3$ and probability of total correct detections $ Pd $  vs. time $T_3 $  in a time constraint $T_1+T_2+T_3=1$ such that $ (T_1,T_2,T_3) := (\frac{1-T_3 }{2},\frac{1-T_3 }{2},T_3 )$ where $0 \le T_3 \le 1$. \todo[disable,inline]{\texttt{\detokenize{Gauss_MonteCarlo_MI_Pd_T3.m \newline Gauss_MI_2}}                     }	}
	\label{f0}
\end{figure}
\subsection{Vector Gaussian channel}  \label{dermi}
Mutual information between two random vectors can be defined as the difference between the total differential entropy in one random vector and the conditional differential entropy in the second random vector given the first vector. We write
\begin{IEEEeqnarray*}{rCl}
I(X;Y)&=& H(Y)-H(Y|X) \label{e5} \IEEEyesnumber
\end{IEEEeqnarray*}  

The conditional entropy $ H(Y|X) $ is calculated from the conditional probability mass functions $ f(Y|X_i) $ defined as
\begin{IEEEeqnarray*}{rCl}
&&f(Y|X=[\lambda_0 \quad \lambda_0]^{\intercal})\nonumber\\ &&= \> \mathcal{N}(Y_1; \lambda_0 \sqrt{ T_1},1) \cdot \mathcal{N}(Y_2; \lambda_0 \sqrt{T_2},1) \cdot \mathcal{N}(Y_3; 2\lambda_0 \sqrt{T_3},1),  \\
&&f(Y|X=[\lambda_0 \quad \lambda_1]^{\intercal})\nonumber\\ &&= \>\mathcal{N}(Y_1; \lambda_0 \sqrt{T_1},1) \cdot \mathcal{N}(Y_2; \lambda_1 \sqrt{T_2},1) \cdot  \mathcal{N}(Y_3; (\lambda_0 + \lambda_1) \sqrt{T_3},1),  \\
&&f(Y|X=[\lambda_1 \quad \lambda_0]^{\intercal})\nonumber\\ &&= \>\mathcal{N}(Y_1; \lambda_1 \sqrt{T_1},1) \cdot \mathcal{N}(Y_2; \lambda_0 \sqrt{T_2},1) \cdot  \mathcal{N}(Y_3;(\lambda_1+\lambda_0) \sqrt{T_3},1),  \\
&&f(Y|X=[\lambda_1 \quad \lambda_1]^{\intercal}) \nonumber\\ &&= \>\mathcal{N}(Y_1; \lambda_1 \sqrt{T_1},1) \cdot \mathcal{N}(Y_2; \lambda_1 \sqrt{T_2},1) \cdot  \mathcal{N}(Y_3; 2\lambda_1 \sqrt{T_3},1). 
\end{IEEEeqnarray*}
The marginal probability mass function of $Y$ is then given as
	\begin{IEEEeqnarray*}{rCl}
	&&f(Y)\nonumber\\ &&= (1-p)^2 \cdot f(Y|X=[\lambda_0 \quad \lambda_0]^{\intercal}) + p(1-p) \cdot \nonumber\\ && f(Y|X=[\lambda_0 \quad \lambda_1]^{\intercal})  +  p(1-p) \cdot f(Y|X=[\lambda_1 \quad \lambda_0]^{\intercal})  + \nonumber\\ && p^2 \cdot f(Y|X=[\lambda_1 \quad \lambda_1]^{\intercal}). \label{e05} \IEEEyesnumber
	\end{IEEEeqnarray*}
As illustrated in fig.(\ref{Figg_6}) the sensing paradigm for our problem; mutual information $I(X;Y)$ is then defined as
\begin{IEEEeqnarray*}{rCl}
	I(X;Y)=H(Y)- H(Y|X), \label{AL0} \IEEEyesnumber
\end{IEEEeqnarray*}
where $H(Y)$ is a differential entropy of our finite Gaussian mixture model (gmm) and given as
\begin{IEEEeqnarray*}{rCl}
	H(Y)	&=&  - \int_{-\infty}^{\infty} \int_{-\infty}^{\infty} \int_{-\infty}^{\infty}\: \Bigg[ \Bigg( (1-p)^2 \cdot  \mathcal{N}(y_1;\lambda_0 T_1, 1) \cdot \mathcal{N}(y_2; \lambda_0 T_2,1) \cdot  \mathcal{N}(y_3; 2\lambda_0 T_3,1)  + \\ && \> p(1-p) \cdot \mathcal{N}(y_1; \lambda_0 T_1,1) \cdot \mathcal{N}(y_2; \lambda_1 T_2,1) \cdot  \mathcal{N}(y_3; (\lambda_0+\lambda_1) T_3,1)   + \\ && \> p(1-p) \cdot  \mathcal{N}(y_1; \lambda_1 T_1,1) \cdot \mathcal{N}(y_2; \lambda_0 T_2,1) \cdot  \mathcal{N}(y_3; (\lambda_1+\lambda_0) T_3,1)   + \\ && \> p^2 \cdot \mathcal{N}(y_1; \lambda_1 T_1,1) \cdot \mathcal{N}(y_2; \lambda_1 T_2,1) \cdot  \mathcal{N}(y_3; 2 \lambda_1 T_3,1) \Bigg)  \cdot \\ && \> \Bigg( \operatorname{Log}_2[ (1-p)^2 \cdot  \mathcal{N}(y_1;\lambda_0 T_1,1) \cdot \mathcal{N}(y_2; \lambda_0 T_2,1) \cdot  \mathcal{N}(y_3; 2\lambda_0 T_3,1)  + \\ && \> p(1-p) \cdot \mathcal{N}(y_1; \lambda_0 T_1,1) \cdot \mathcal{N}(y_2; \lambda_1 T_2, 1) \cdot  \mathcal{N}(y_3; (\lambda_0+\lambda_1) T_3,1)   + \\ && \> p(1-p) \cdot  \mathcal{N}(y_1; \lambda_1 T_1,1) \cdot \mathcal{N}(y_2; \lambda_0 T_2,1) \cdot  \mathcal{N}(y_3; (\lambda_1+\lambda_0) T_3,1)   + \\ && \> p^2 \cdot \mathcal{N}(y_1; \lambda_1 T_1,1) \cdot \mathcal{N}(y_2; \lambda_1 T_2,1) \cdot  \mathcal{N}(y_3; 2 \lambda_1 T_3,1)  ] \Bigg) \Bigg] \,d{y_1}\,d{y_2}\,d{y_3}, \label{AL2} \IEEEyesnumber
\end{IEEEeqnarray*}
and $H(Y|X)$ is 
\begin{IEEEeqnarray*}{rCl}
	H(Y|X)	&=& (1-p)^2\cdot (0.5 \cdot \operatorname{Log}_2[2 \pi \: e ] + 0.5 \cdot \operatorname{Log}_2[2 \pi \: e]  + 0.5 \cdot \operatorname{Log}_2[2 \pi \: e ])  + \\ && \>   p(1-p) \cdot (0.5 \cdot \operatorname{Log}_2[2 \pi \: e ] + 0.5 \cdot \operatorname{Log}_2[2 \pi \: e ]  + 0.5 \cdot \operatorname{Log}_2[2 \pi \: e)  + \\ && \>  p(1-p)\cdot (0.5 \cdot \operatorname{Log}_2[2 \pi \: e ] + 0.5 \cdot \operatorname{Log}_2[2 \pi \: e]  + 0.5 \cdot \operatorname{Log}_2[2 \pi \: e ])   \\ && \> p^2\cdot (0.5 \cdot \operatorname{Log}_2[2 \pi \: e] + 0.5 \cdot \operatorname{Log}_2[2 \pi \: e]  + 0.5 \cdot \operatorname{Log}_2[2 \pi \: e ]). \label{AL3} \IEEEyesnumber
\end{IEEEeqnarray*}

Since multidimensional integral defined in (\ref{AL2}) have no closed-form solution, we have to resort to numerical methods.
We may mitigate the curse of dimensionality involved in  multi-dimensional integration by Monte-Carlo technique by taking samples from the multivariate Gaussian mixture distribution to achieve fast convergence to the true mixture differential entropy at a reasonable computational burden; whereas naive uniform sampling of the space would lead to a quite slow convergence to the true differential entropy.
\begin{IEEEeqnarray*}{rCl}
H(Y)&=&E[-\operatorname{Log}_2[f_Y(Y)]]\\ &=& \>-\int_{-\infty}^{\infty} \int_{-\infty}^{\infty} \int_{-\infty}^{\infty} f_Y(y) \cdot \operatorname{Log}_2[f_Y(y)] \,d{y} \\ &\approx& \>-\frac{\sum_i \operatorname{Log}_2[f_Y(s_i)]}{N_s},
\label{AL4} \IEEEyesnumber
\end{IEEEeqnarray*} 	  
where $f_Y(\cdot)$ is the mixture probability distribution of $Y$, $N_s$ is the number of MC samples and $s_i$ is the $i^{th}$ sample from multivariate Gaussian mixture distribution  \cite{verdu2019}.
\begin{theorem}
	$I(X_1,X_2;Y_1,Y_2,Y_3)$ is symmetric in variables $ T_1 $ and $ T_2 $.
\end{theorem}
\begin{IEEEproof}
	Mutual information $I(X_1,X_2;Y_1,Y_2,Y_3)$ given in (\ref{AL0}) is invariant under any permutation of variables $ T_1 $ and $ T_2 $. That means interchanging the two variables leaves the expression unchanged.
\end{IEEEproof}
\begin{theorem}
	$I(X_1,X_2;Y_1,Y_2,Y_3)$ is concave in $T_3=0$ plane.
\end{theorem}
\begin{IEEEproof}
	\\ $I(X_1,X_2;Y_1,Y_2,Y_3)\Big|_{T3=0}=I(X_1,X_2;Y_1,Y_2)$ 
	\\By chain rule of mutual information:
	\begin{IEEEeqnarray*}{rCl}
I(X_1,X_2;Y_1,Y_2)=&& \>I(X_1,X_2;Y_1)+I(X_1,X_2;Y_2|Y_1) \\  
		=&& I(X_1;Y_1) + I(X_2;Y_2). \label{e6} \IEEEyesnumber
	\end{IEEEeqnarray*}
	We note in (\ref{e6}) that each $ I(X_i;Y_i) $ is solely a function of $ T_i $ and also concave in it \cite{guo2005mutual}. Since the sum of concave functions is a concave function. Therefore $I(X_1,X_2;Y_1,Y_2)$ is concave in $T_1$ and $ T_2$ when $T_3=0$. This concludes the proof.
\end{IEEEproof}

\begin{theorem}
$I(X_1,X_2;Y_1,Y_2,Y_3)$ is concave in $T_3$ along the line 
$(T_1,T_2,T_3) \allowbreak := \allowbreak (\frac{T-T_3}{2},\allowbreak \frac{T-T_3}{2},T_3)$  parametrized by $ 0 \le T_3 \le T $. 
\end{theorem}
\begin{IEEEproof}
It is noted in \cite[Theorem 5]{payaro2009hessian} that mutual information is a concave function of the squared singular values $(\boldsymbol{\lambda})$ of the precoder matrix $P$ if the first $m^{'}$ eigenvectors of the channel covariance matrix $(\mathbf{R}_{H}=H^\intercal \mathbf{R}_{Z}^{-1} H)$ coincide with the left singular vectors of the precoder $P$ i-e $\mathsf{H}_\lambda I(S;Y) \le 0$ for the signal model $Y=HPS+Z$ where $H \in \mathbb{R}^{n \times p}$ is the channel, $S$ is the input signaling $S \in \mathbb{R}^{m}$, $P$ is a precoder matrix $P \in \mathbb{R}^{p \times m}$ and $Z \in  \mathbb{R}^{n}$ is Gaussian noise independent of the input $S$ and has covariance matrix $\mathbf{R}_Z$.

For our problem: $H= I,$ $\mathbf{R}^{-1}_Z=\Lambda=I,$ $P=\Phi$, $S=X$ and $Z=N.$ The singular value decomposition of $\Phi=U \Sigma V^*$. We have singular matrix
\begin{equation}
\Sigma=\begin{bmatrix}
\frac{\sqrt{T_1+T_2+2T_3-\sqrt{(T_1-T_2)^2+4 T_3^2}}}{\sqrt{2}} & 0  \\
0 & \frac{\sqrt{T_1+T_2+2T_3+\sqrt{(T_1-T_2)^2+4 T_3^2}}}{\sqrt{2}} \\
0 & 0 \\
\end{bmatrix}
\end{equation}
By substituting $T_1=T_2=\frac{T-T_3}{2}$, the squared singular values are $[ \lambda_1 \quad \lambda_2 \quad \lambda_3]= \allowbreak [ \frac{T-T_3}{2} \quad \frac{T+3 \cdot T_3}{2} \quad 0  ]$ for $0 \le T_3 \le T.$  This is just the composition with an affine transformation on the domain. Concavity remains preserved under affine transformation \cite[page 79-86 ]{boyd2004convex}.
\end{IEEEproof}
The above concavity of $I$ is illustrated in fig.(\ref{f0}).
\section{Detection Theoretic Description} \label{detd}
\subsection{Bayes risk} \label{adoptcri}
In last section, we presented the metric of mutual information $ I $ between hidden random vector $ X, $ and observable vector $ Y $. Here we approach the sensing problem as a multi-hypothesis detection problem and define the Bayesian risk \cite[pp.220]{schonhoff2006detection} to minimize in $(T_1,T_2,T_3)$. We define Bayes risk $ r $ as
\begin{IEEEeqnarray*}{rCl}
	r &=& (1-p)^2 \Big [P_{\lambda_0 \lambda_0 \:| \: \lambda_0 \lambda_0} \: C_{\lambda_0 \lambda_0 \:| \: \lambda_0 \lambda_0} + P_{\lambda_0 \lambda_1 \:| \: \lambda_0 \lambda_0} \: C_{\lambda_0 \lambda_1 \:| \: \lambda_0 \lambda_0}  \nonumber\\ && \>+P_{\lambda_1 \lambda_0 \:| \: \lambda_0 \lambda_0} \: C_{\lambda_1 \lambda_0 \:| \: \lambda_0 \lambda_0}+ P_{\lambda_1 \lambda_1 \:| \: \lambda_0 \lambda_0} \: C_{\lambda_1 \lambda_1 \:| \: \lambda_0 \lambda_0} \Big] + p(1-p)\nonumber\\ && \>\Big [P_{\lambda_0 \lambda_0 \:| \: \lambda_0 \lambda_1} \: C_{\lambda_0 \lambda_0 \:| \: \lambda_0 \lambda_1} + P_{\lambda_0 \lambda_1 \:| \: \lambda_0 \lambda_1} \: C_{\lambda_0 \lambda_1 \:| \: \lambda_0 \lambda_1} + P_{\lambda_1 \lambda_0 \:| \: \lambda_0 \lambda_1} \: C_{\lambda_1 \lambda_0 \:| \: \lambda_0 \lambda_1}\nonumber\\ && \>+ P_{\lambda_1 \lambda_1 \:| \: \lambda_0 \lambda_1} \: C_{\lambda_1 \lambda_1 \:| \: \lambda_0 \lambda_1} \Big]+  p(1-p) \Big [P_{\lambda_0 \lambda_0 \:| \: \lambda_1 \lambda_0} \: C_{\lambda_0 \lambda_0 \:| \: \lambda_1 \lambda_0} \nonumber\\ && \>+ P_{\lambda_0 \lambda_1 \:| \: \lambda_1 \lambda_0} \: C_{\lambda_0 \lambda_1 \:| \: \lambda_1 \lambda_0} + P_{\lambda_1 \lambda_0 \:| \: \lambda_1 \lambda_0} \: C_{\lambda_1 \lambda_0 \:| \: \lambda_1 \lambda_0}+ P_{\lambda_1 \lambda_1 \:| \: \lambda_1 \lambda_0} \: C_{\lambda_1 \lambda_1 \:| \: \lambda_1 \lambda_0} \Big] \nonumber\\ && \> +p^2 \Big [P_{\lambda_0 \lambda_0 \:| \: \lambda_1 \lambda_1} \: C_{\lambda_0 \lambda_0 \:| \: \lambda_1 \lambda_1} + P_{\lambda_0 \lambda_1 \:| \: \lambda_1 \lambda_1} \: C_{\lambda_0 \lambda_1 \:| \: \lambda_1 \lambda_1} + P_{\lambda_1 \lambda_0 \:| \: \lambda_1 \lambda_1} \: C_{\lambda_1 \lambda_0 \:| \: \lambda_1 \lambda_1}\nonumber\\ && \>+ P_{\lambda_1 \lambda_1 \:| \: \lambda_1 \lambda_1} \: C_{\lambda_1 \lambda_1 \:| \: \lambda_1 \lambda_1} \Big],
\end{IEEEeqnarray*}
where $ P_{\lambda_k \lambda_l \:| \: \lambda_i \lambda_j}  $ is the probability that $ X=[\lambda_i,\lambda_j]^{\intercal} $ is true while decision $ X=[\lambda_k,\lambda_l]^{\intercal}$ is made; similarly for $ C_{\lambda_k \lambda_l \:| \: \lambda_i \lambda_j} $. Setting all costs for which $ [\lambda_i,\lambda_j]^{\intercal} \neq [\lambda_k,\lambda_l]^{\intercal} $ to one and  $[\lambda_i,\lambda_j]^{\intercal} = [\lambda_k,\lambda_l]^{\intercal}  $ to zero, we have 
\begin{IEEEeqnarray}{rCl}
	r &=& (1-p)^2 \Big [P_{\lambda_0 \lambda_1 \:| \: \lambda_0 \lambda_0} \: + P_{\lambda_1 \lambda_0 \:| \: \lambda_0 \lambda_0} \: + P_{\lambda_1 \lambda_1 \:| \: \lambda_0 \lambda_0} \: \Big] +  p(1-p) \nonumber\\ && \>\Big [P_{\lambda_0 \lambda_0 \:| \: \lambda_0 \lambda_1} \:  +  P_{\lambda_1 \lambda_0 \:| \: \lambda_0 \lambda_1} \:+ P_{\lambda_1 \lambda_1 \:| \: \lambda_0 \lambda_1} \:  \Big]+ p(1-p)\nonumber\\ && \> \Big [P_{\lambda_0 \lambda_0 \:| \: \lambda_1 \lambda_0} \:  + P_{\lambda_0 \lambda_1 \:| \: \lambda_1 \lambda_0} \: + P_{\lambda_1 \lambda_1 \:| \: \lambda_1 \lambda_0} \:  \Big]+   p^2\nonumber\\ && \> \Big [P_{\lambda_0 \lambda_0 \:| \: \lambda_1 \lambda_1} \:  + P_{\lambda_0 \lambda_1 \:| \: \lambda_1 \lambda_1} \:  + P_{\lambda_1 \lambda_0 \:| \: \lambda_1 \lambda_1} \:  \Big]. \label{e2}
\end{IEEEeqnarray}  
We are interested in minimizing this Bayes risk $ r $ in $ (T_1,T_2,T_3)$ i-e
\begin{IEEEeqnarray*}{rCl}
	& \underset{T_1, T_2, T_3}{\text{min}} \: r \quad \text{s.t.} \: \:  T_1+T_2+T_3=T. \label{e4} \IEEEyesnumber
\end{IEEEeqnarray*} 
Note that while minimizing $ r $ in $ (T_1,T_2,T_3)$, the decisions boundaries would be changing accordingly and become function of $ (T_1,T_2,T_3)$.
Equivalently, we may say that
\begin{IEEEeqnarray*}{rCl}
	& \underset{T_1, T_2, T_3}{\text{max}} \: P_d \quad  \text{s.t.} \: \:  T_1+T_2+T_3=T \label{e8} \IEEEyesnumber
\end{IEEEeqnarray*}
where $ P_d $ is probability of total correct detections, $ P_d=1-r. $ In the next section we present the computed results of (\ref{e8}).
\section{Monte Carlo Simulation Results} \label{comand}
\begin{figure*}[ht]
	\begin{subfigure}{.49\textwidth}
		\includegraphics[width=\linewidth]{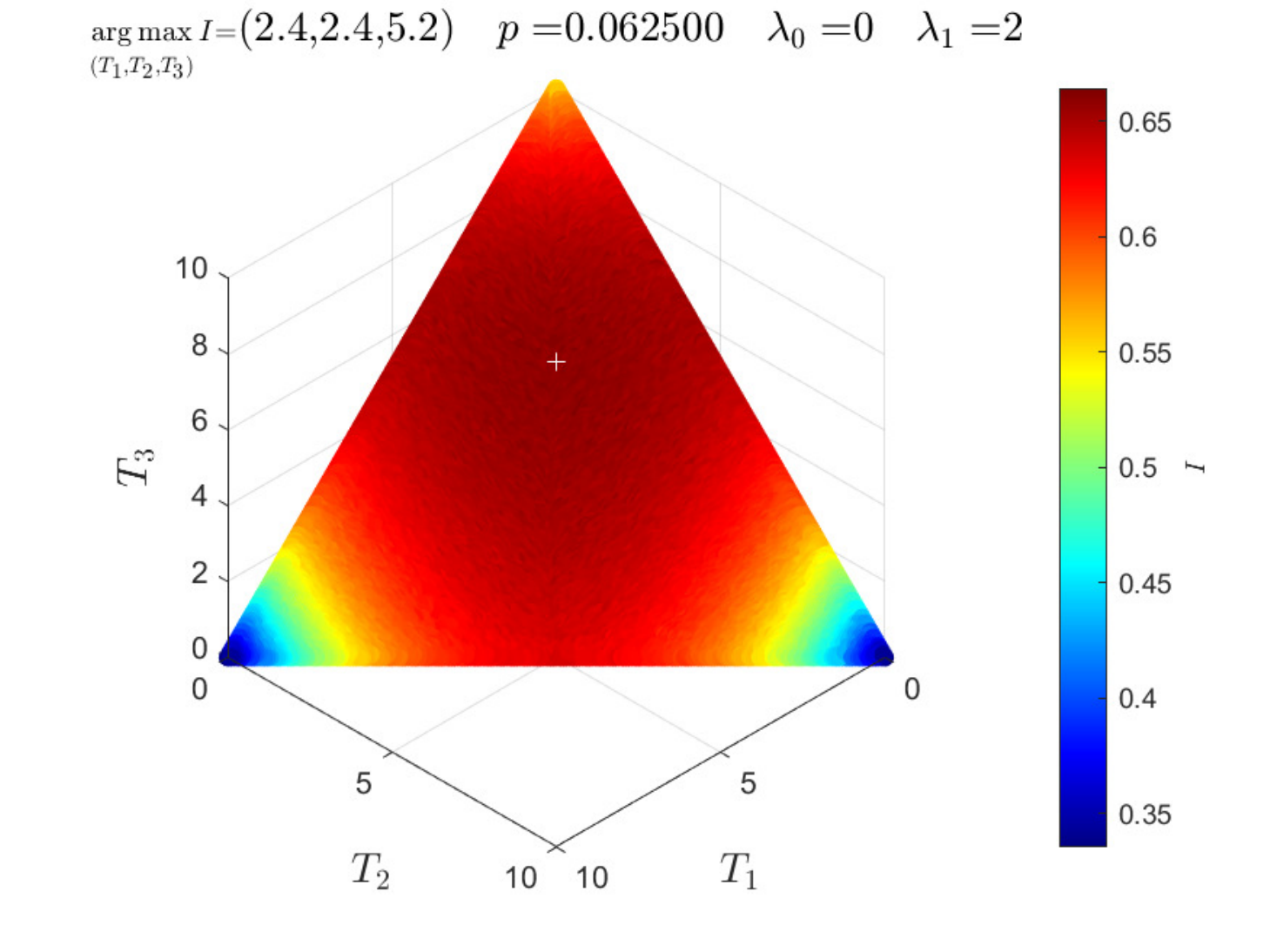}
		\caption{ }
		\label{fig2a}
	\end{subfigure} 
	\begin{subfigure}{.49\textwidth}
		\includegraphics[width=\linewidth]{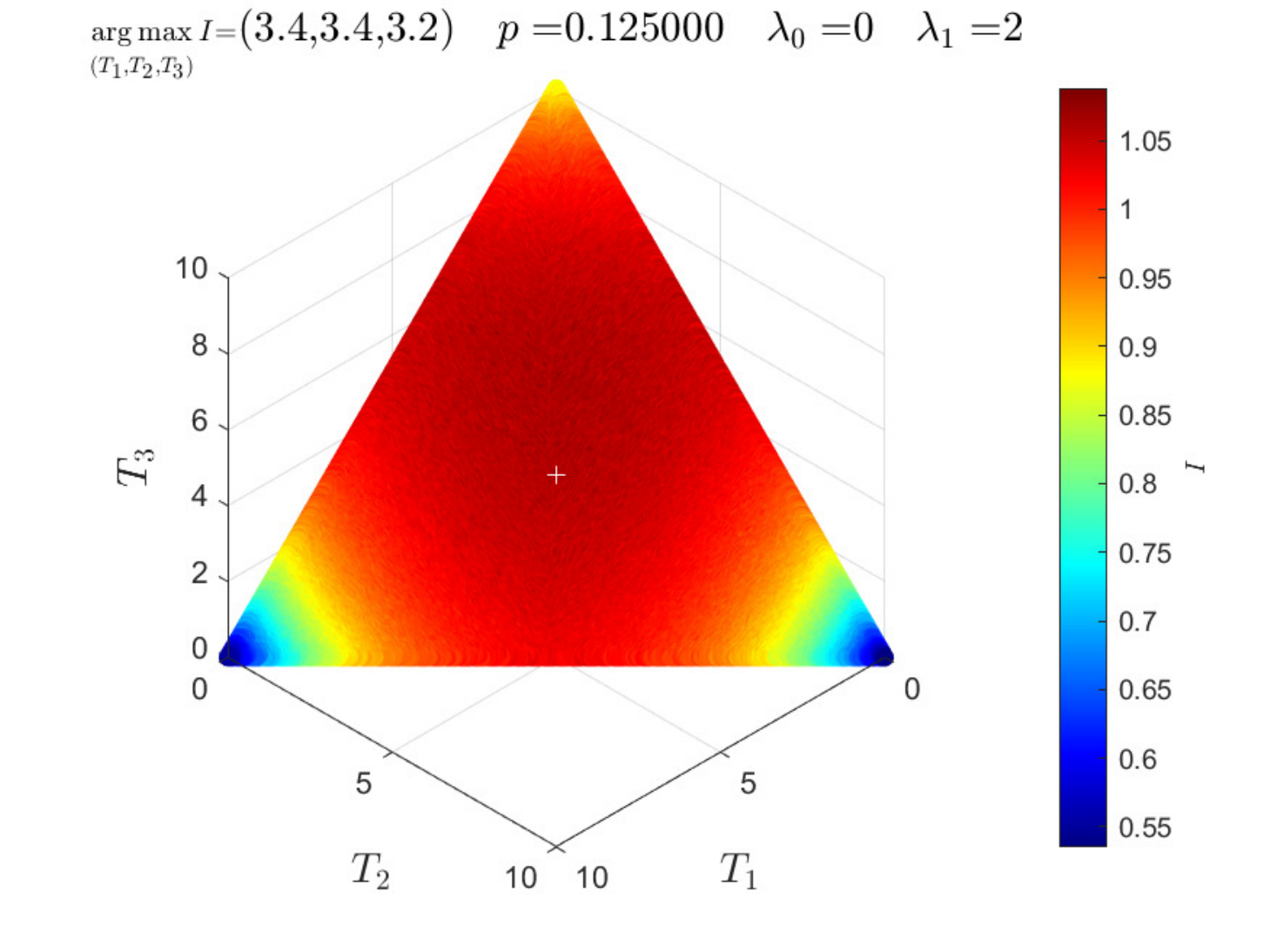}
		\caption{ }
		\label{fig2b}
	\end{subfigure} \\%
	\begin{subfigure}{.49\textwidth}
		\includegraphics[width=\linewidth]{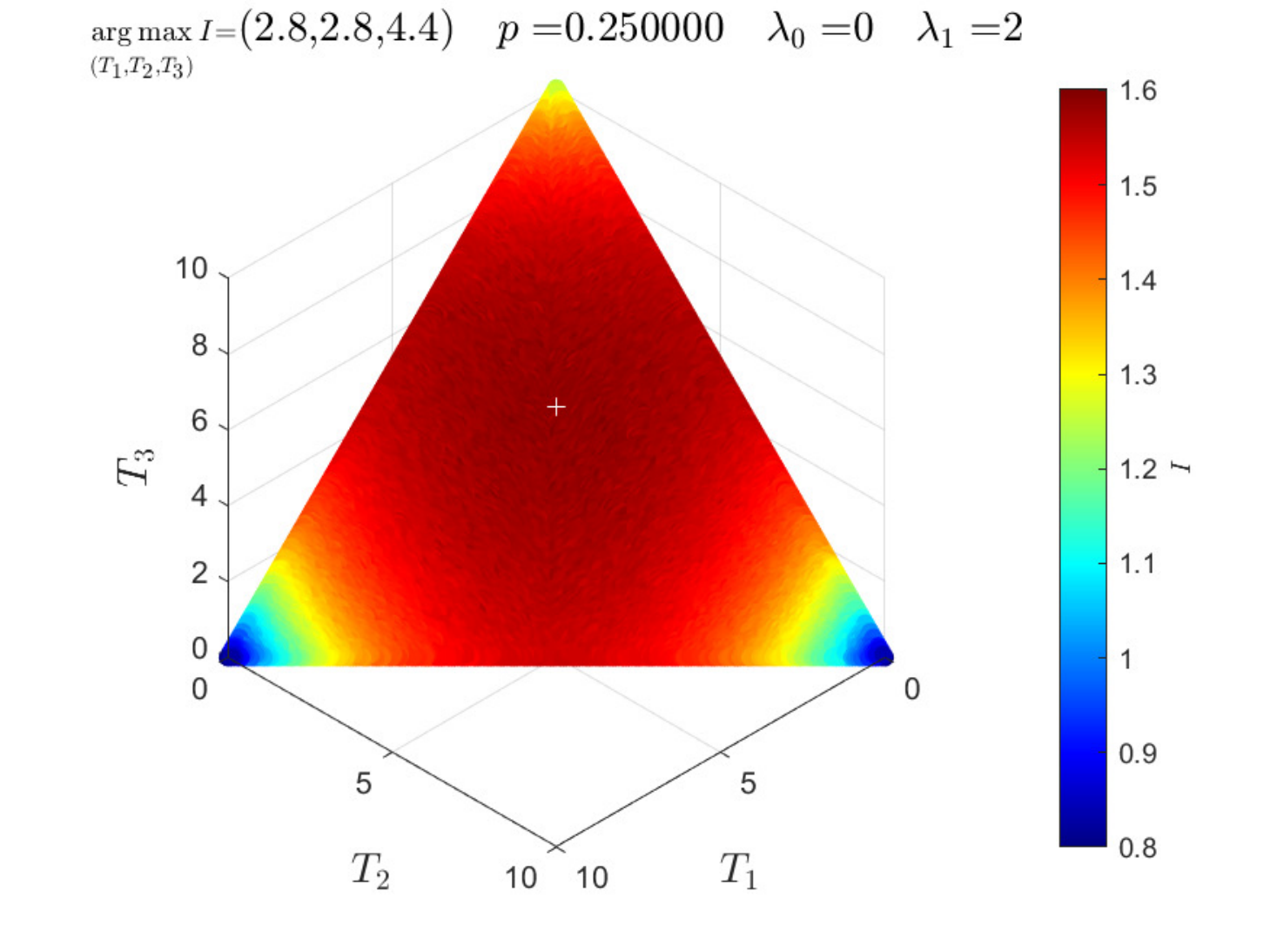}
		\caption{ }
		\label{fig2c}
	\end{subfigure} %
	\begin{subfigure}{.49\textwidth}
		\includegraphics[width=\linewidth]{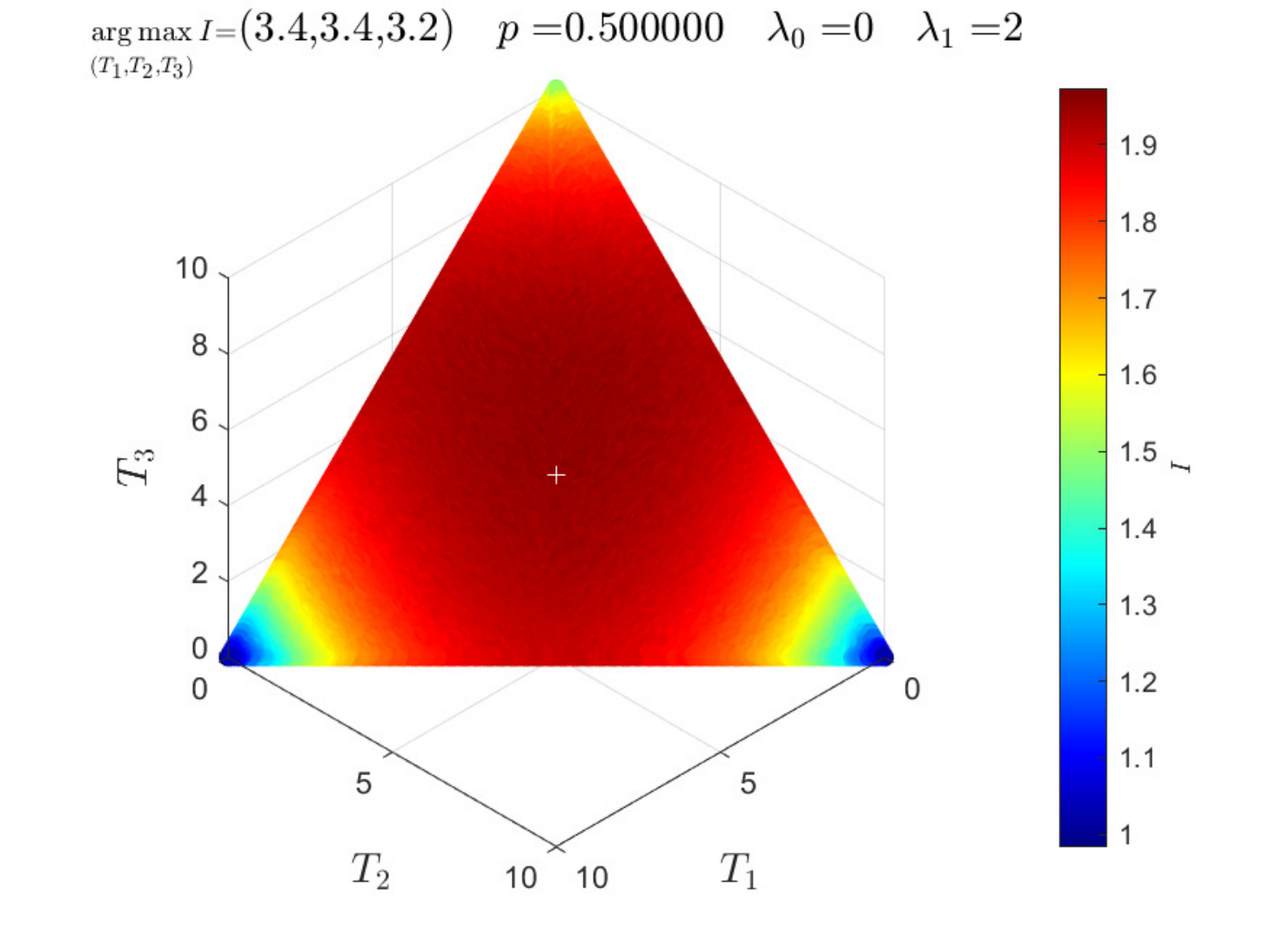}
		\caption{ }
		\label{fig2d}
	\end{subfigure}
		\begin{subfigure}{.49\textwidth}
			\includegraphics[width=\linewidth]{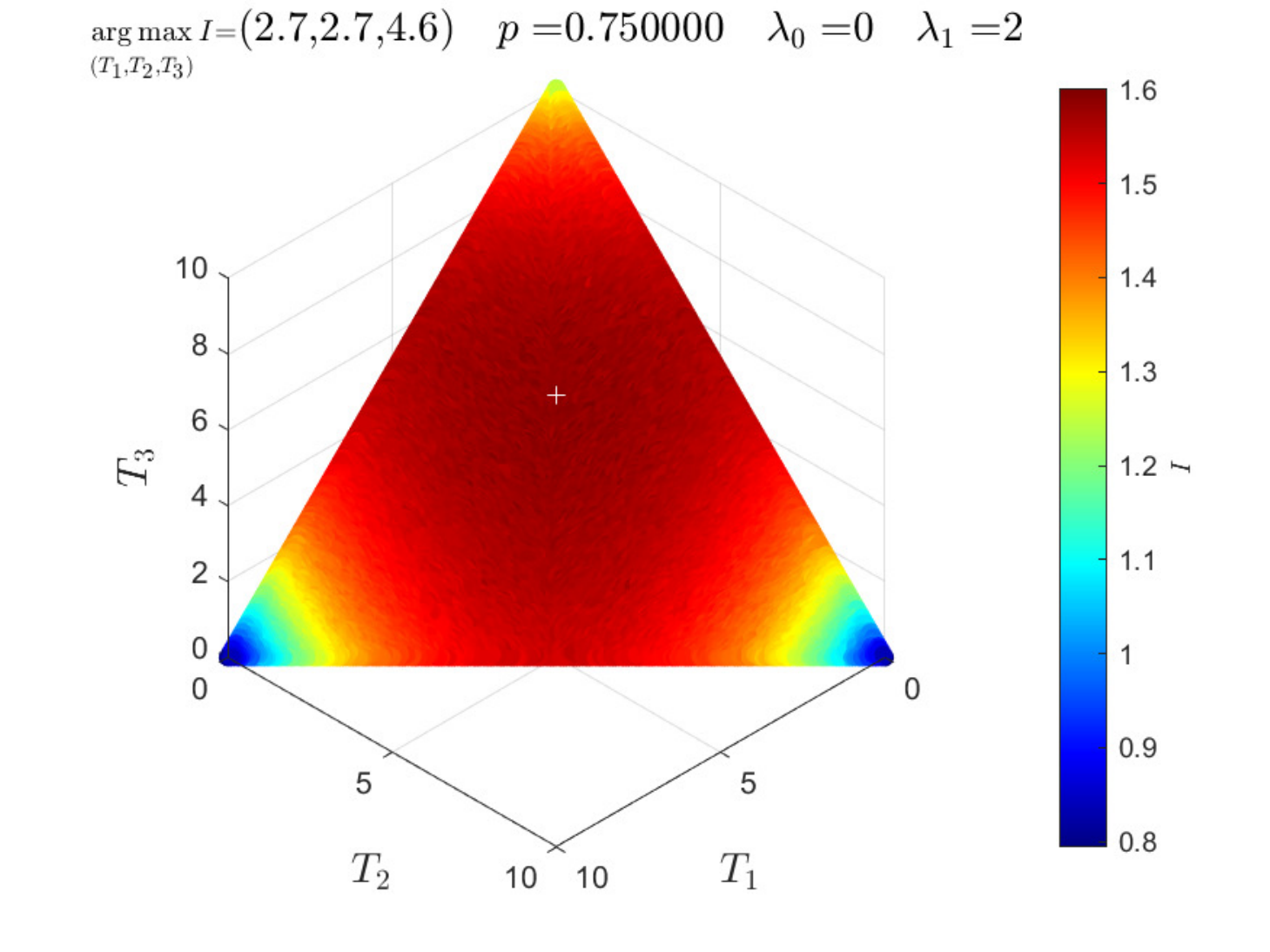}
			\caption{ }
			\label{fig2e}
		\end{subfigure} %
			\begin{subfigure}{.49\textwidth}
				\includegraphics[width=\linewidth]{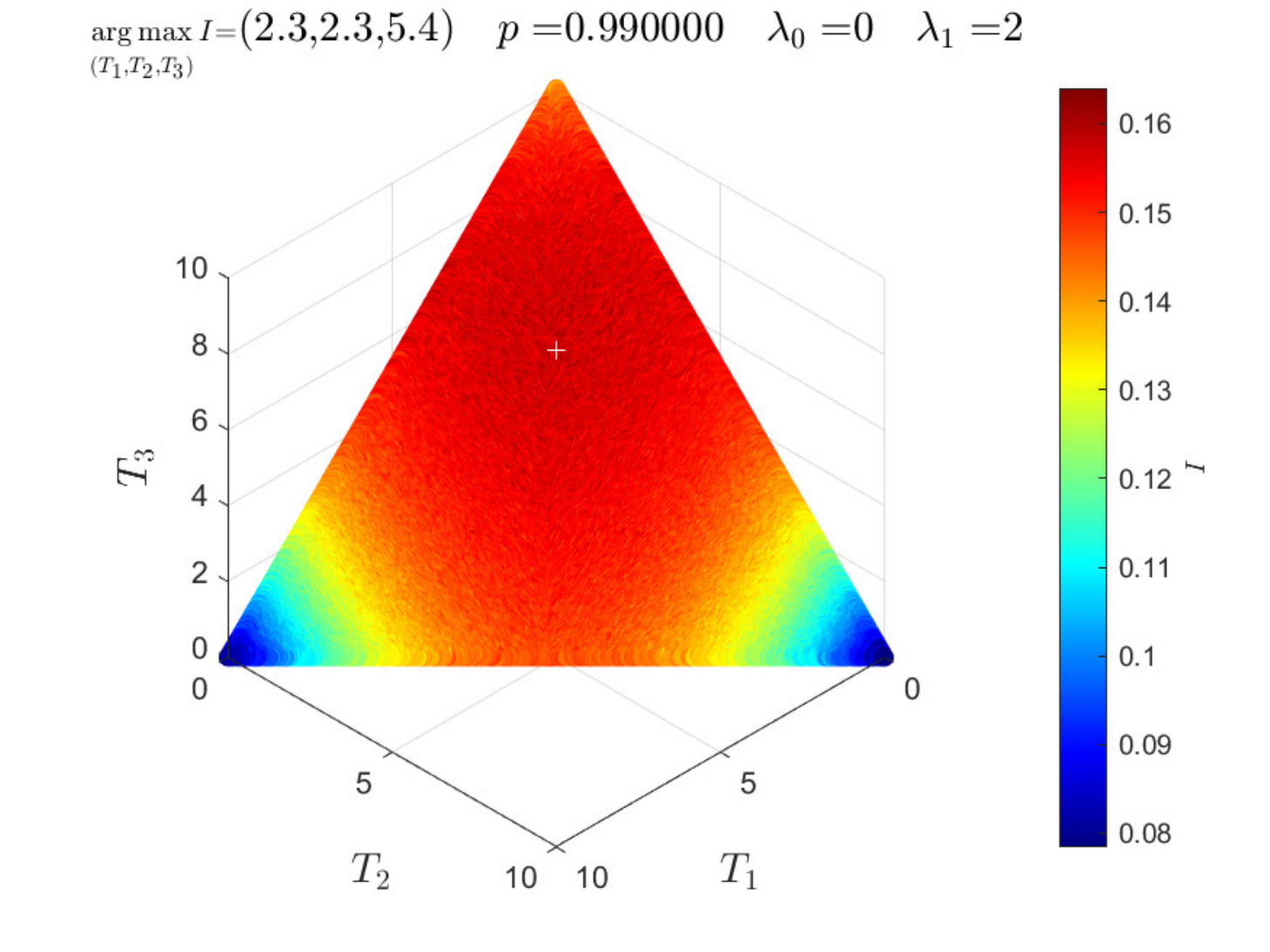}
				\caption{ }
				\label{fig2f}
			\end{subfigure} %
	\caption{$ I(X;Y) $ vs. $ (T_1,T_2,T_3) $ under time constraint $ T_1+T_2+T_3=10 $ for $ \lambda_0=0 $, $ \lambda_1=2 $,  and varying \emph{prior} probability $ p $.  \todo[disable,inline]{\texttt{\detokenize{Gaussian_Two_bin.m}}
			\newline   	\texttt{\detokenize{Gauss_MI.m}}                     }}
	\label{f2}
\end{figure*}

For all simulation purposes, we have assumed that optimizing argument in $ \underset{T_1, T_2, T_3}{\text{max}} \allowbreak \quad I(X_1, X_2 ; Y_1, Y_2, Y_3) $ $\text{s.t.}$  $T_1+T_2+T_3=T$  would have $T_1=T_2$. This is based on the observations noted in the ternary diagrams given in  fig.(\ref{f2}). We computed $I$ for a wide range of given input parameters $\lambda_0$, $\lambda_1$, $T$ and $p$; and it was noted that maximizing argument always seems to lie on the line $ (T_1,T_2,T_3) := (\frac{T-\alpha}{2},\frac{T-\alpha}{2},\alpha)$ where $0 \le \alpha \le T$. In other words we noted a Schur concavity of $I$ in $(T1,T2)$ whenever $T_3$ is held fixed under a given time-constraint; however no proof of Schur concavity of $I$ is claimed in this work.

We compute mutual information in (\ref{AL2}) by first computing the entropy of the multivariate Gaussian mixture by generating the samples from it. Each of the Gaussian mixture component is a $3-$dimensional multivariate Gaussian distribution that comes with a prior belief. We generate a total of $10^6$ samples to calculate $H(Y)$ for a given prior $p$ and energy constraint $T_1+T_2+T_3=T$ with $T_1=T_2$. Since we do have a closed-form available for a differential entropy of a multivariate Gaussian distribution therefore for $H(Y|X)$ we don't need to apply the Monte Carlo method for evaluating it. The difference of the two would provide the approximated value of $I(X;Y) \Big|_{(\frac{T-\alpha}{2},\frac{T-\alpha}{2},\alpha)}$ for a given set of parameters.

For the MAP detection we use the empirical method to calculate the probability of total correct detections $P_d$. We again assume that the optimal solution has $T_1=T_2$. An optimal solution for any given set of parameters is then searched in the region $ (T_1,T_2,T_3) := (\frac{T-\alpha}{2},\frac{T-\alpha}{2},\alpha)$ where $0 \le \alpha \le T$. For a given value of $T$: $\alpha$ takes $400$ linear steps from $0$ to $T$ and for each step we first generate the samples from the Gaussian mixture under consideration by additionally knowing which mixture component has actually generated any particular sample. For every input sample we then computed the posterior probability for each of four hypotheses and then decide in favor of the hypothesis that has the highest posterior probability. Comparing our $10^6$ decisions with that of the $10^6$ inputs, we can then calculate the total correct detections for each of the discrete $\alpha$ steps. This way for any given set of parameters $\lambda_0$, $\lambda_1$, $T$, $p$ and $\alpha$ we may empirically compute the $P_d\Big|_{(\frac{T-\alpha}{2},\frac{T-\alpha}{2},\alpha)}$.

\begin{figure}[ht]
	\centering
	\begin{subfigure}{.49\textwidth}
	\includegraphics[width=\linewidth]{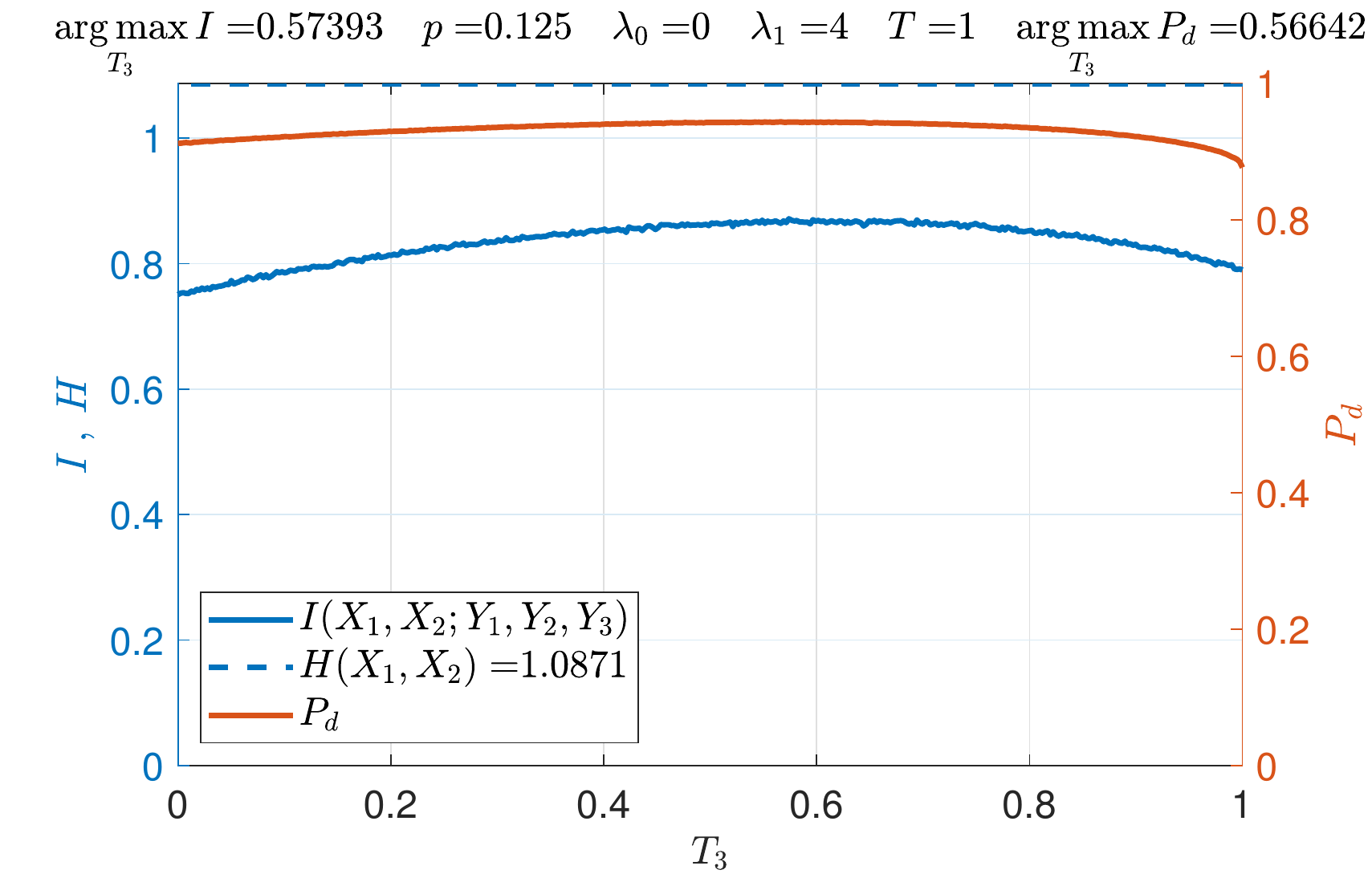}
		\caption{}
		\label{fig3a}
	\end{subfigure}
	\begin{subfigure}{.49\textwidth}
	\includegraphics[width=\linewidth]{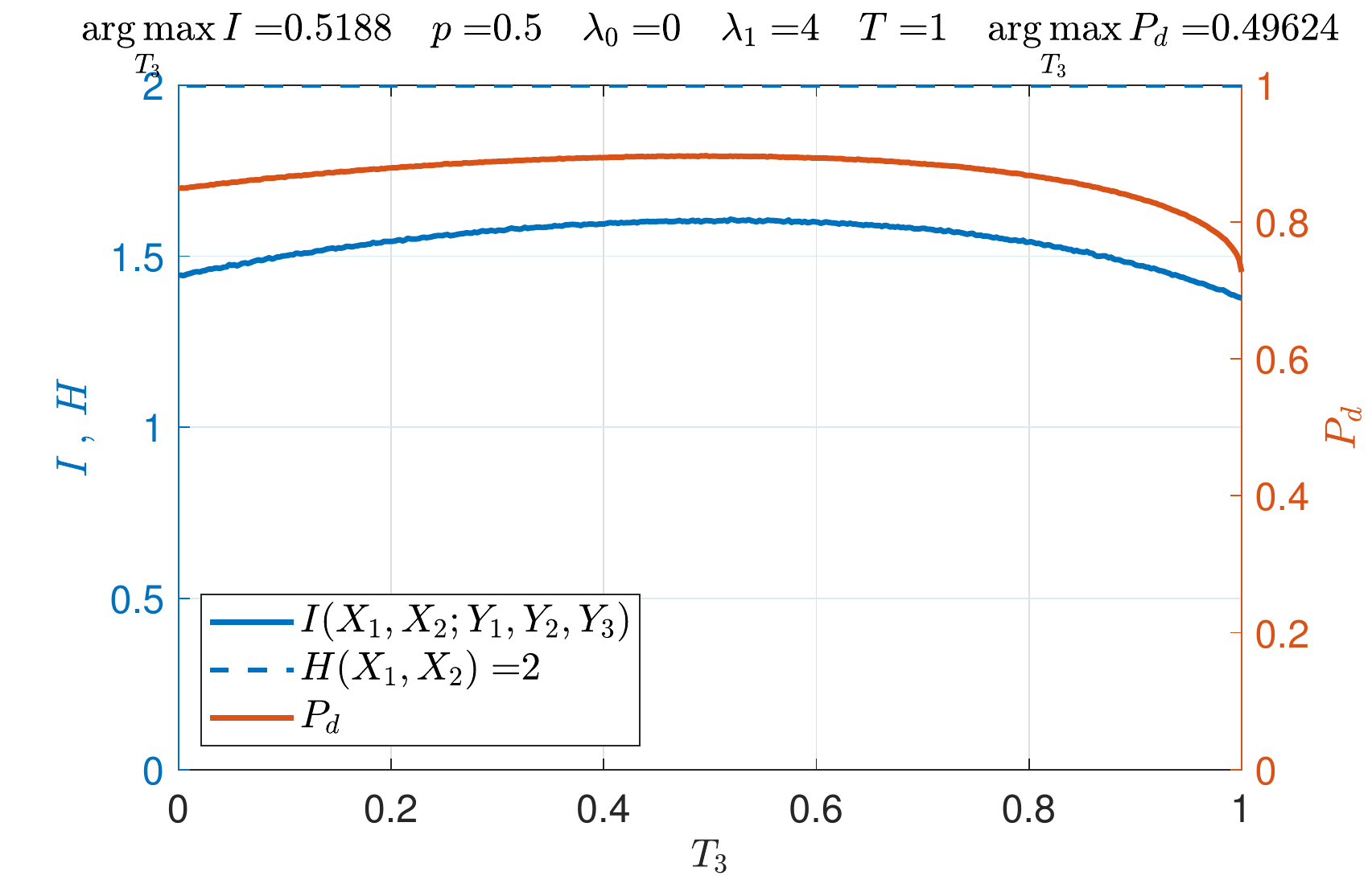}
		\caption{}
		\label{fig3b}
	\end{subfigure} %
	\begin{subfigure}{.49\textwidth}
	\includegraphics[width=\linewidth]{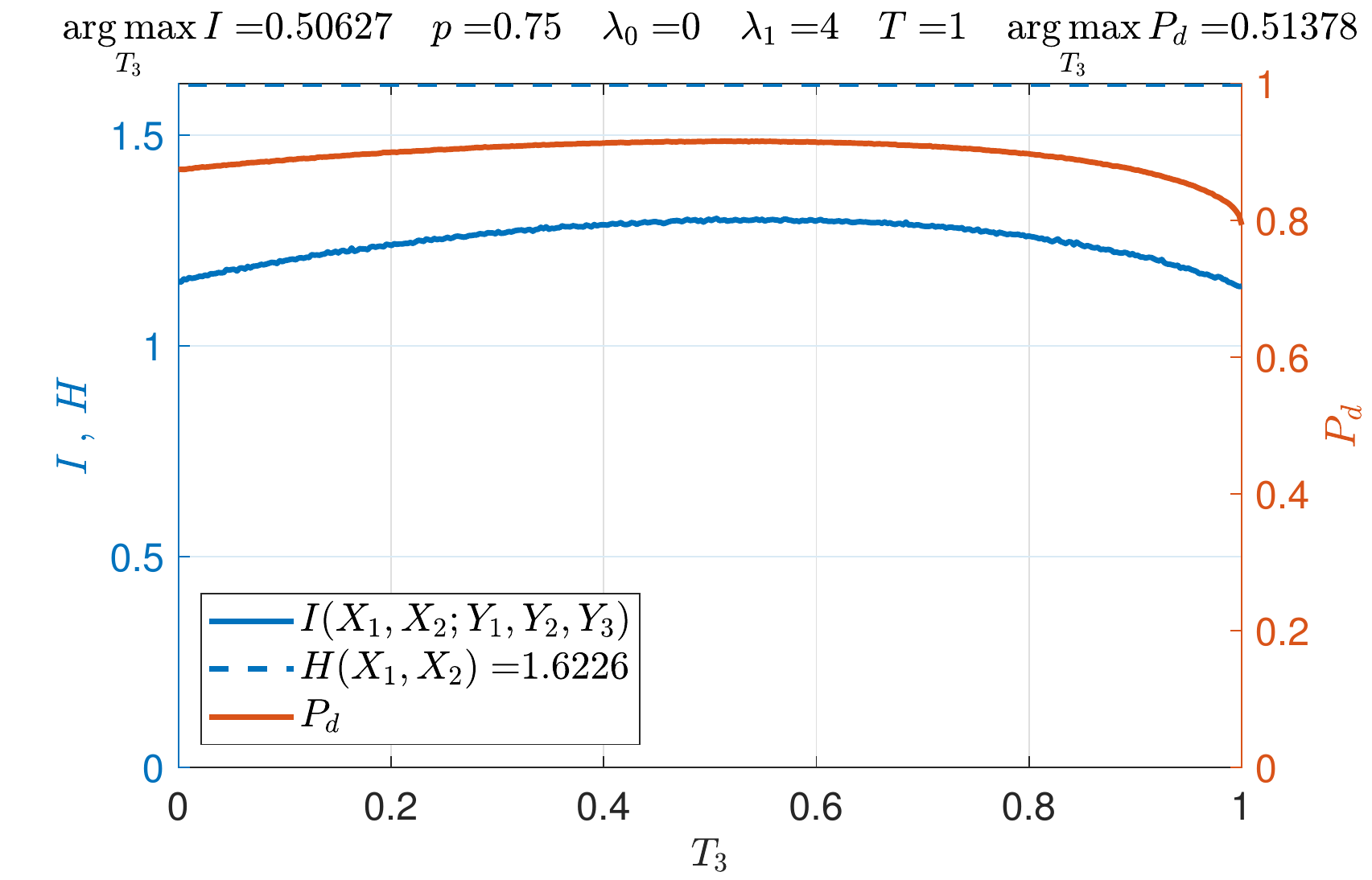}
		\caption{}
		\label{fig3c}
	\end{subfigure} %
	\begin{subfigure}{.49\textwidth}
	\includegraphics[width=\linewidth]{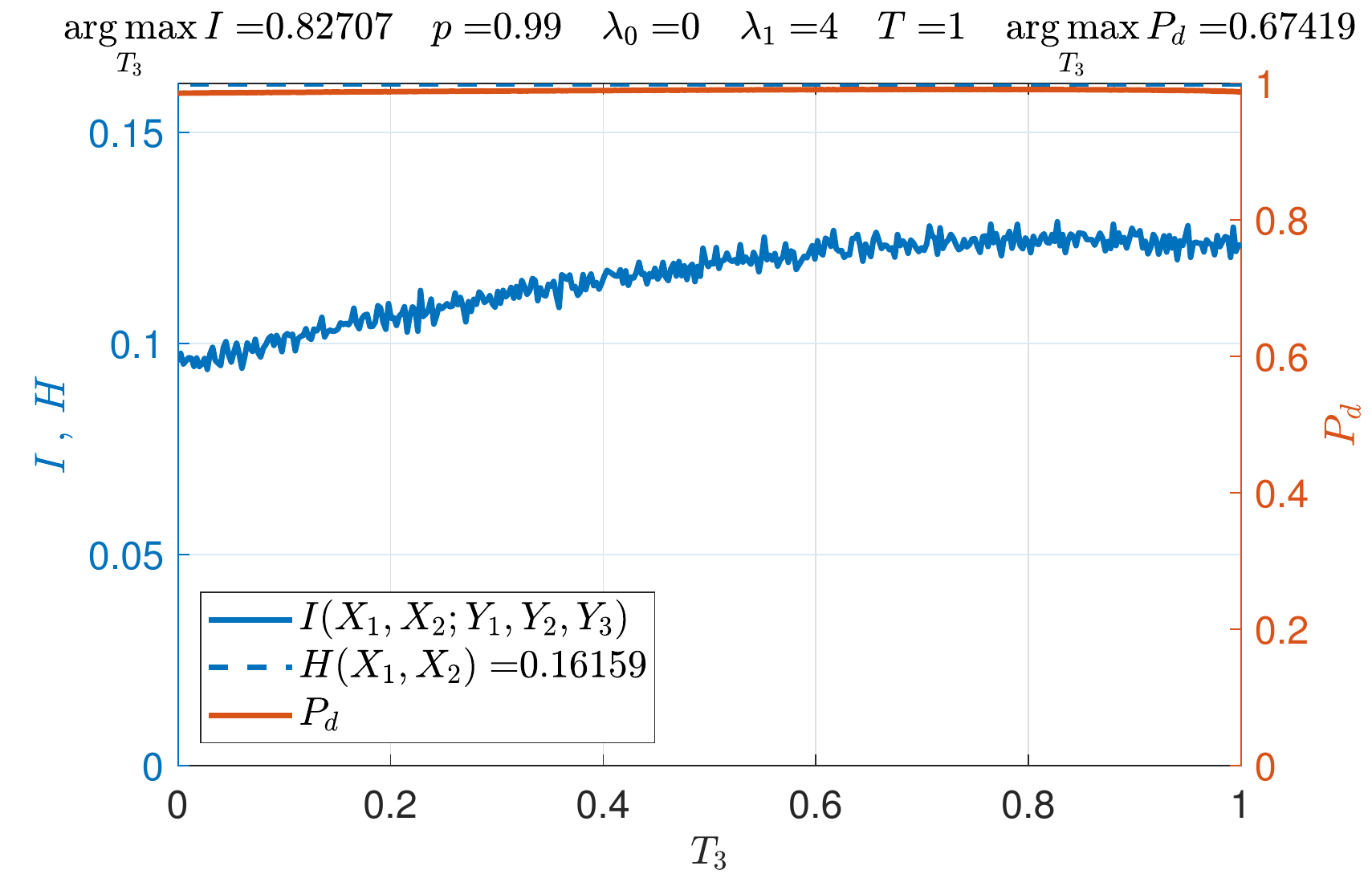}
		\caption{}
		\label{fig3d}
	\end{subfigure}
\caption{Mutual information $ I(X;Y) $ vs. $ T_3 $ and probability of total correct detections $P_d$ vs. $ T_3 $ for \emph{prior} probabilities of $ 0.125,0.5,0.75 $ and $ 0.99. $ 
\todo[disable,inline]{Gauss\_MonteCarlo\_MI\_Pd\_T3.m 
	\newline   	Gauss\_MI\_2.m            }}
\label{f3} 
\end{figure}

\begin{figure}[ht]
	\centering
	\begin{subfigure}{.49\textwidth}
		\includegraphics[width=\linewidth]{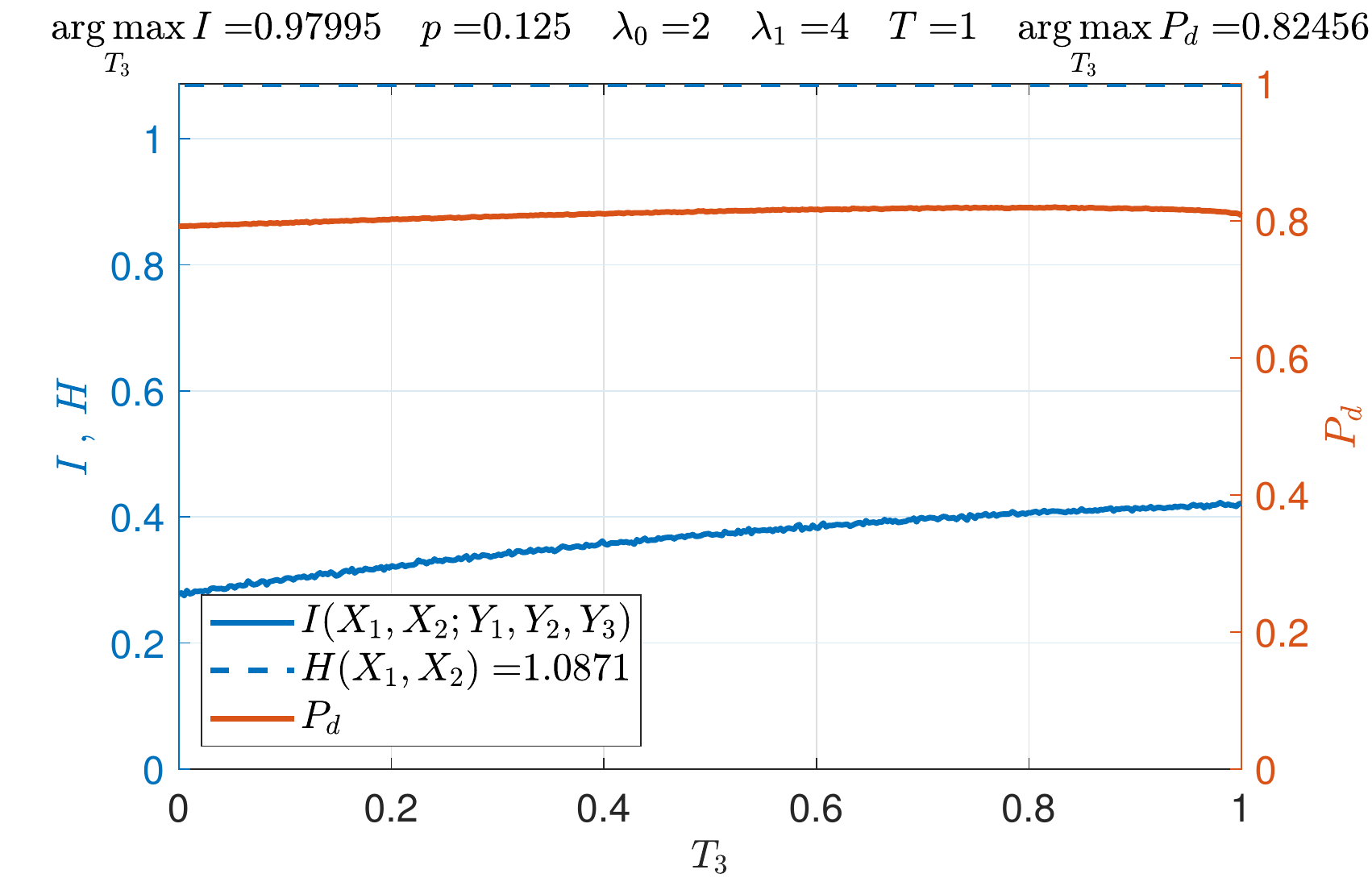}
		\caption{}
		\label{fig9a}
	\end{subfigure}
	\begin{subfigure}{.49\textwidth}
		\includegraphics[width=\linewidth]{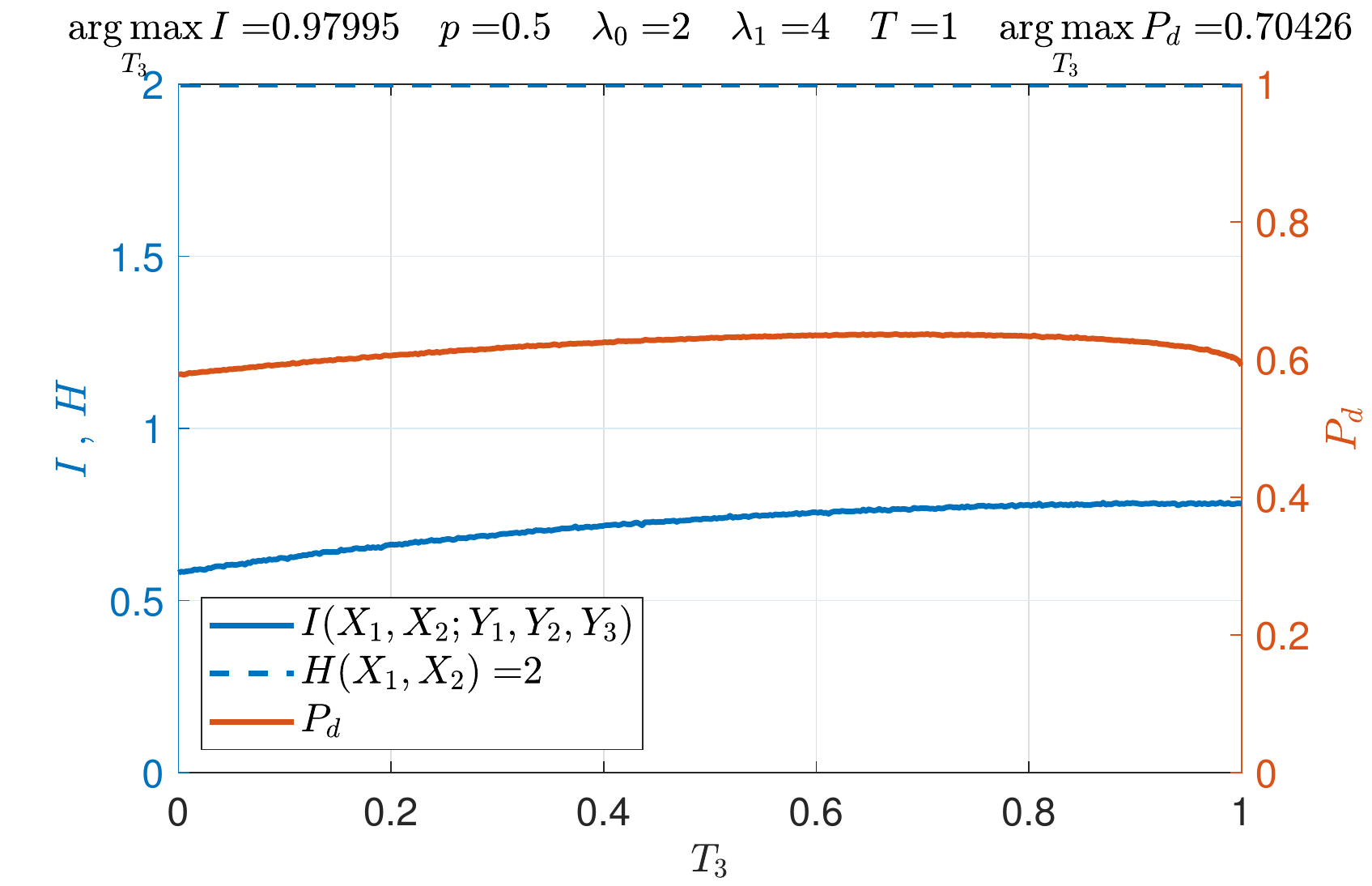}
		\caption{}
		\label{fig9b}
	\end{subfigure} %
	\begin{subfigure}{.49\textwidth}
		\includegraphics[width=\linewidth]{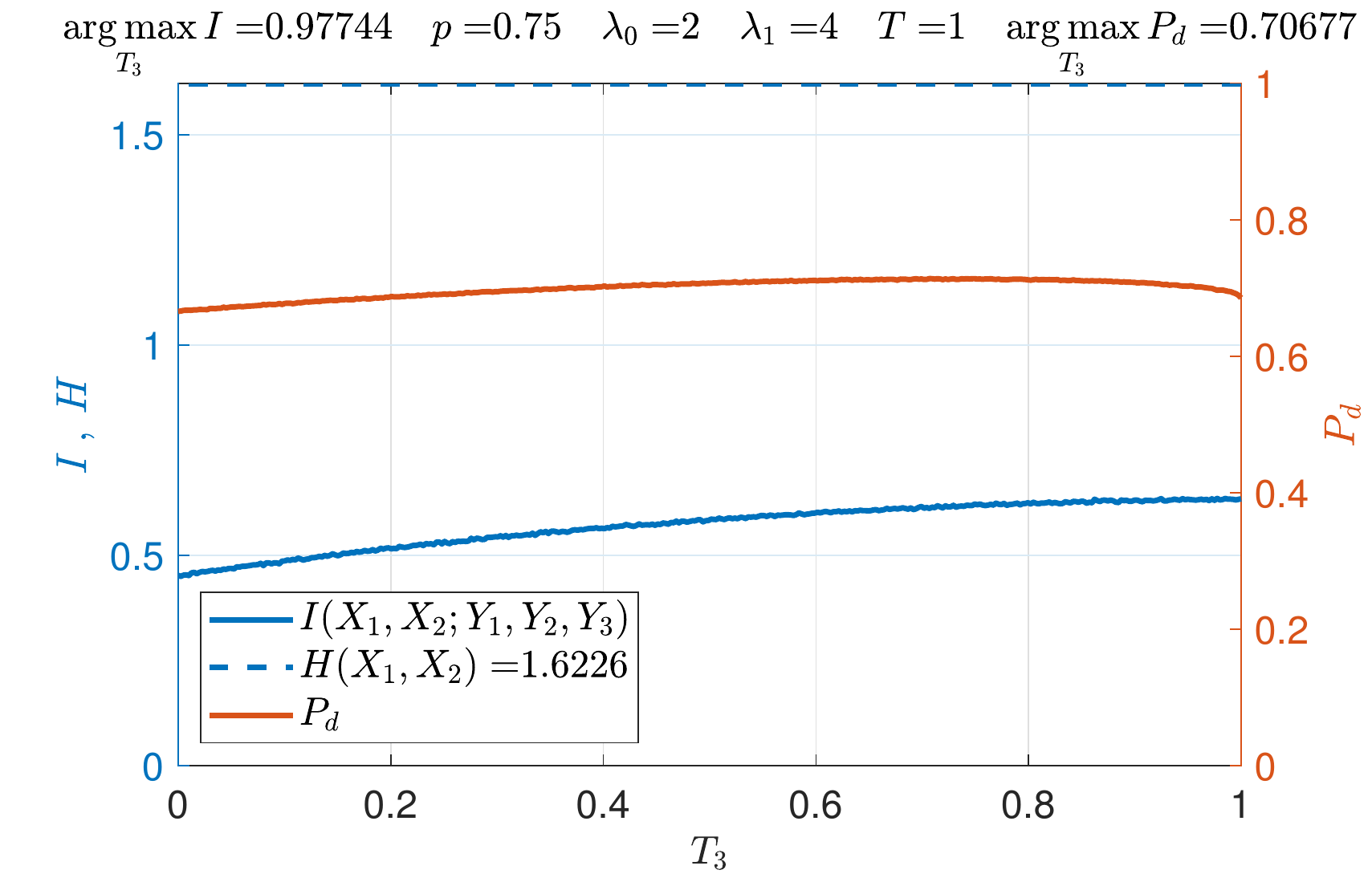}
		\caption{}
		\label{fig9c}
	\end{subfigure} %
	\begin{subfigure}{.49\textwidth}
		\includegraphics[width=\linewidth]{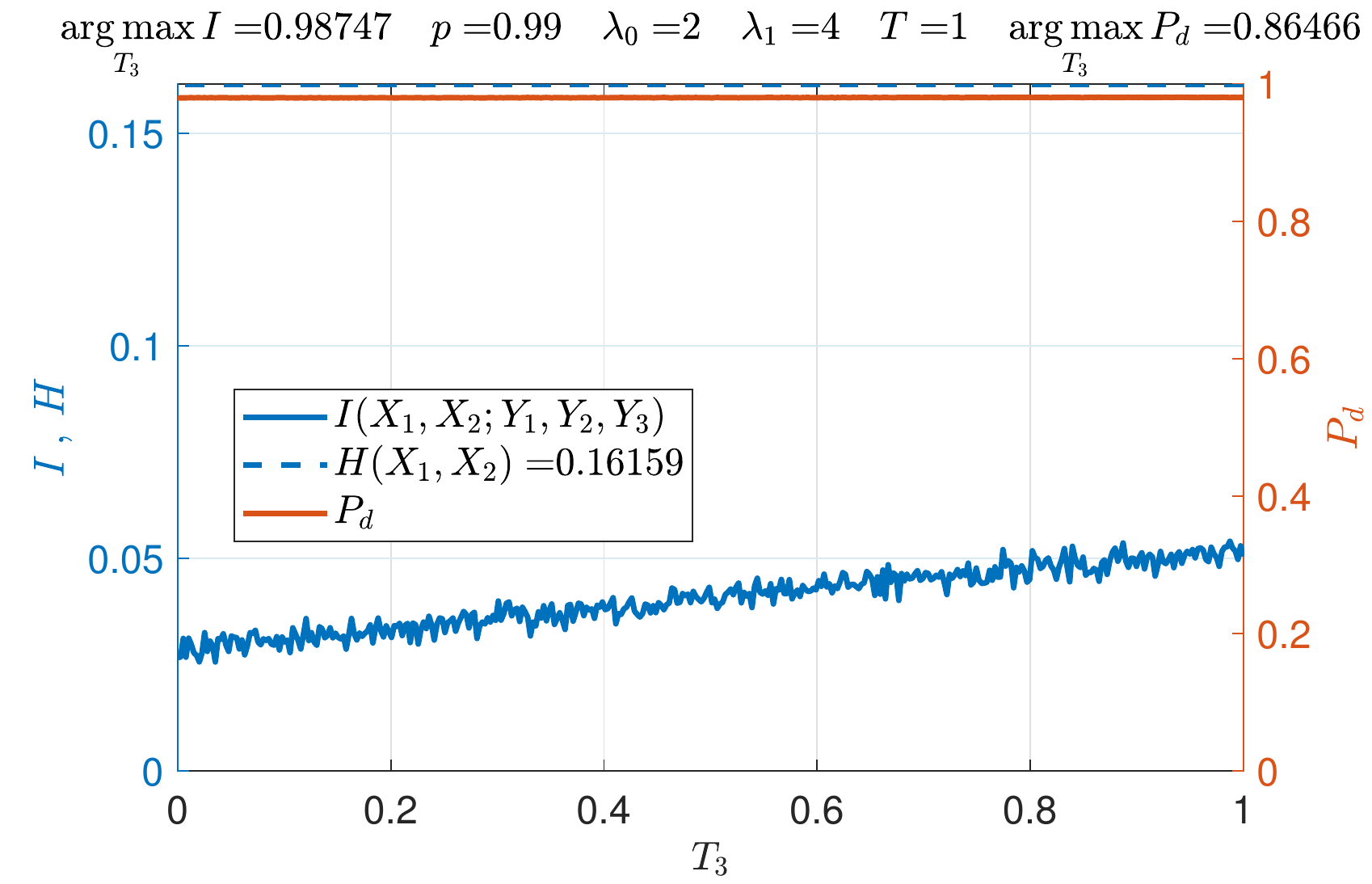}
		\caption{}
		\label{fig9d}
	\end{subfigure}
	\caption{Mutual information $ I(X;Y) $ vs. $ T_3 $ and probability of total correct detections $P_d$ vs. $ T_3 $ for \emph{prior} probabilities of $ 0.125,0.5,0.75 $ and $ 0.99. $ 
		\todo[disable,inline]{\texttt{\detokenize{Gauss_MonteCarlo_MI_Pd_T3.m }}
			\newline   	\texttt{\detokenize{Gauss_MI_2.m}}                     }}
	\label{f9} 
\end{figure}

For the sake of simplicity we call the time proportion: $(\frac{T}{2},\frac{T}{2},0)$ to be the individual sensing; $(0,0,T)$ to be the joint sensing and $(\frac{T-\alpha}{2},\frac{T-\alpha}{2},\alpha)$ where $0 < \alpha < T$ to be hybrid sensing method. In fig.(\ref{f3}) and fig.(\ref{f9}), there are a couple of observations to be noted: first we can see the concavity of $I$ and $P_d$ in $T_3$; second observation is maximizing the mutual information and probability of total correct detections doesn't lead to the same optimal solution; this is more noticeable in fig.(\ref{f9}) where mutual information is maximizing in the vicinity of $T_3=1$ and therefore suggesting individual sensing to be optimum whereas probability of total correct detections is suggesting the hybrid sensing to be optimum. The third observation is that just by looking at the prior $p$ we can't say in the most rough sense that which of the three sensing mechanisms would be optimal, either from the perspective of the mutual information or from the Bayes inference. This is unlike to that of a Poisson problem in \cite{paper1} where individual sensing was always optimal whenever $p \ge 0.5$ irrespective of the given input set of parameters from mutual information perspective.

\begin{figure}[ht]
	\begin{subfigure}{.49\textwidth}
		\includegraphics[width=\linewidth]{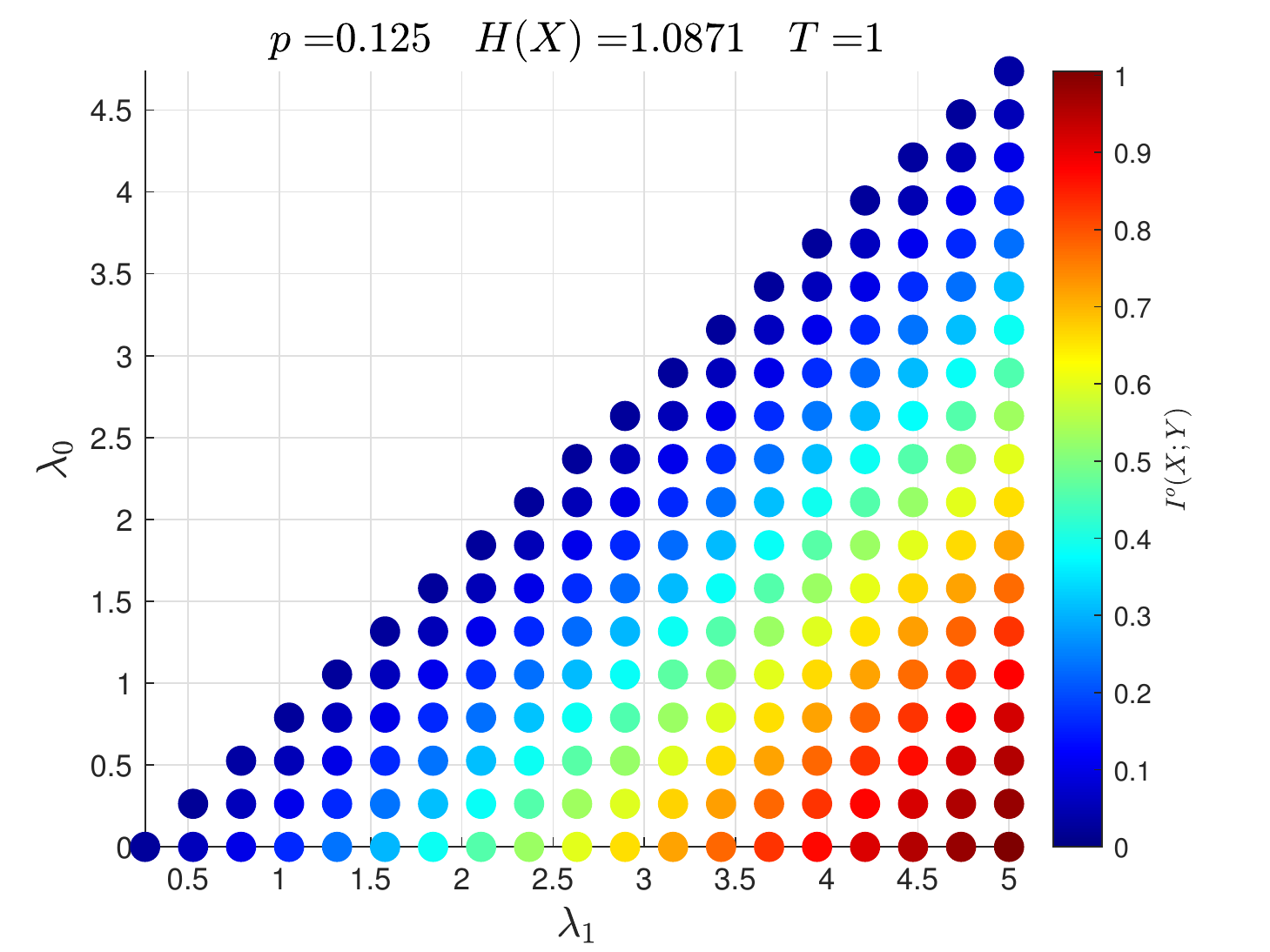}
		\caption{ }
		\label{fig6a}
	\end{subfigure} 
	\begin{subfigure}{.49\textwidth}
		\includegraphics[width=\linewidth]{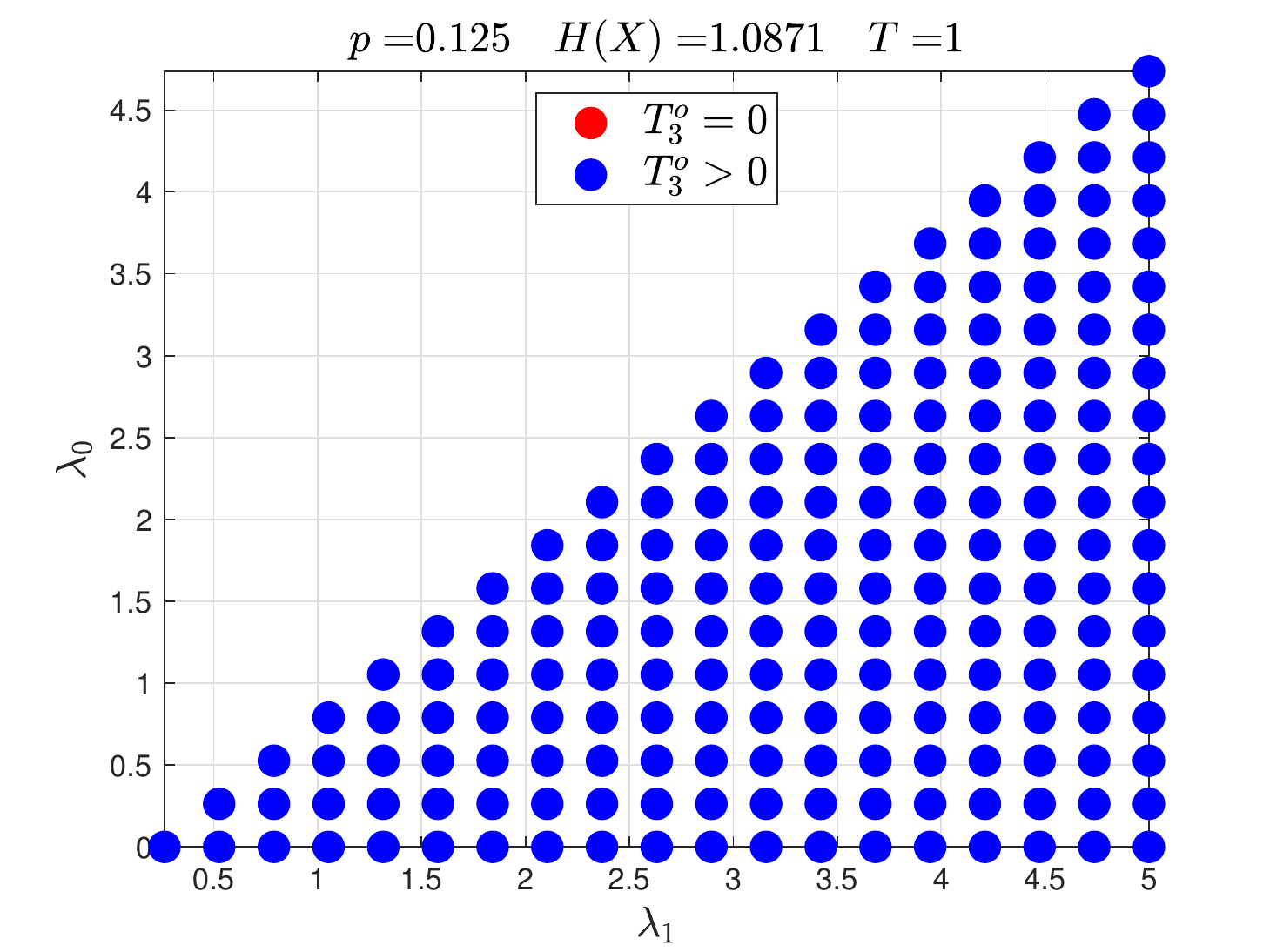}
		\caption{ }
		\label{fig6b}
	\end{subfigure} \\%
	\begin{subfigure}{.49\textwidth}
		\includegraphics[width=\linewidth]{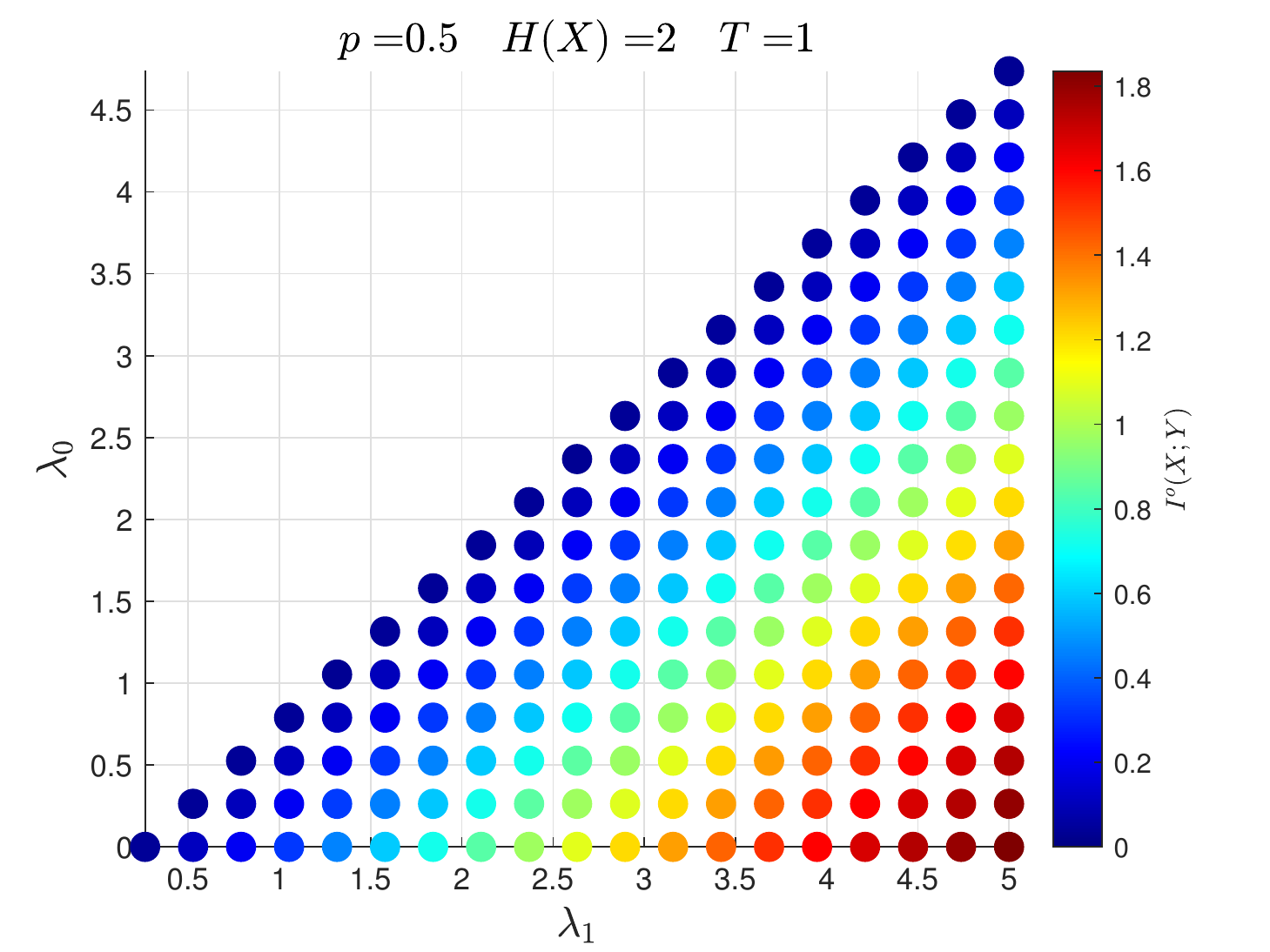}
		\caption{ }
		\label{fig6c}
	\end{subfigure} %
	\begin{subfigure}{.49\textwidth}
		\includegraphics[width=\linewidth]{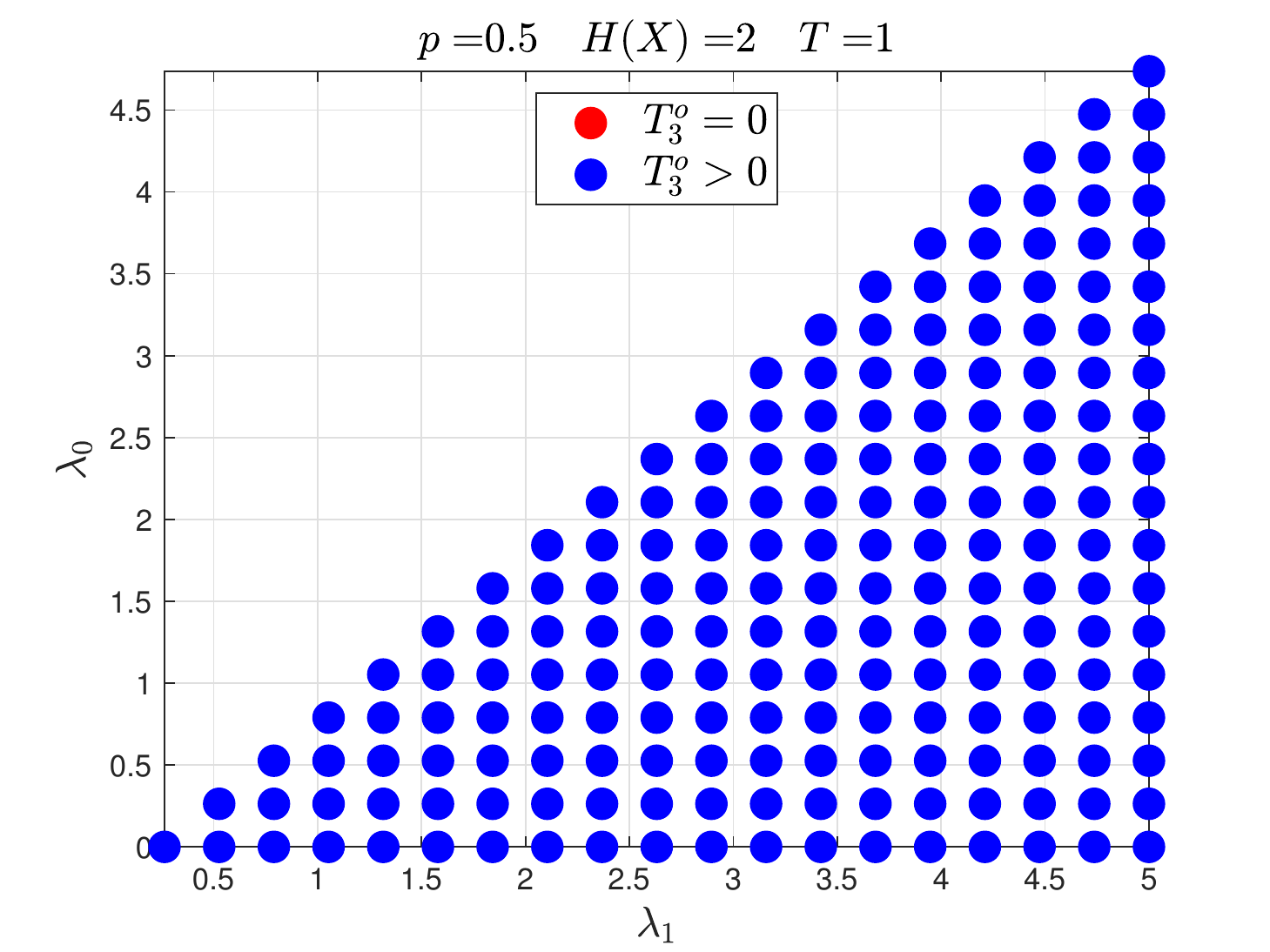}
		\caption{ }
		\label{fig6d}
	\end{subfigure}
	\begin{subfigure}{.49\textwidth}
 		\includegraphics[width=\linewidth]{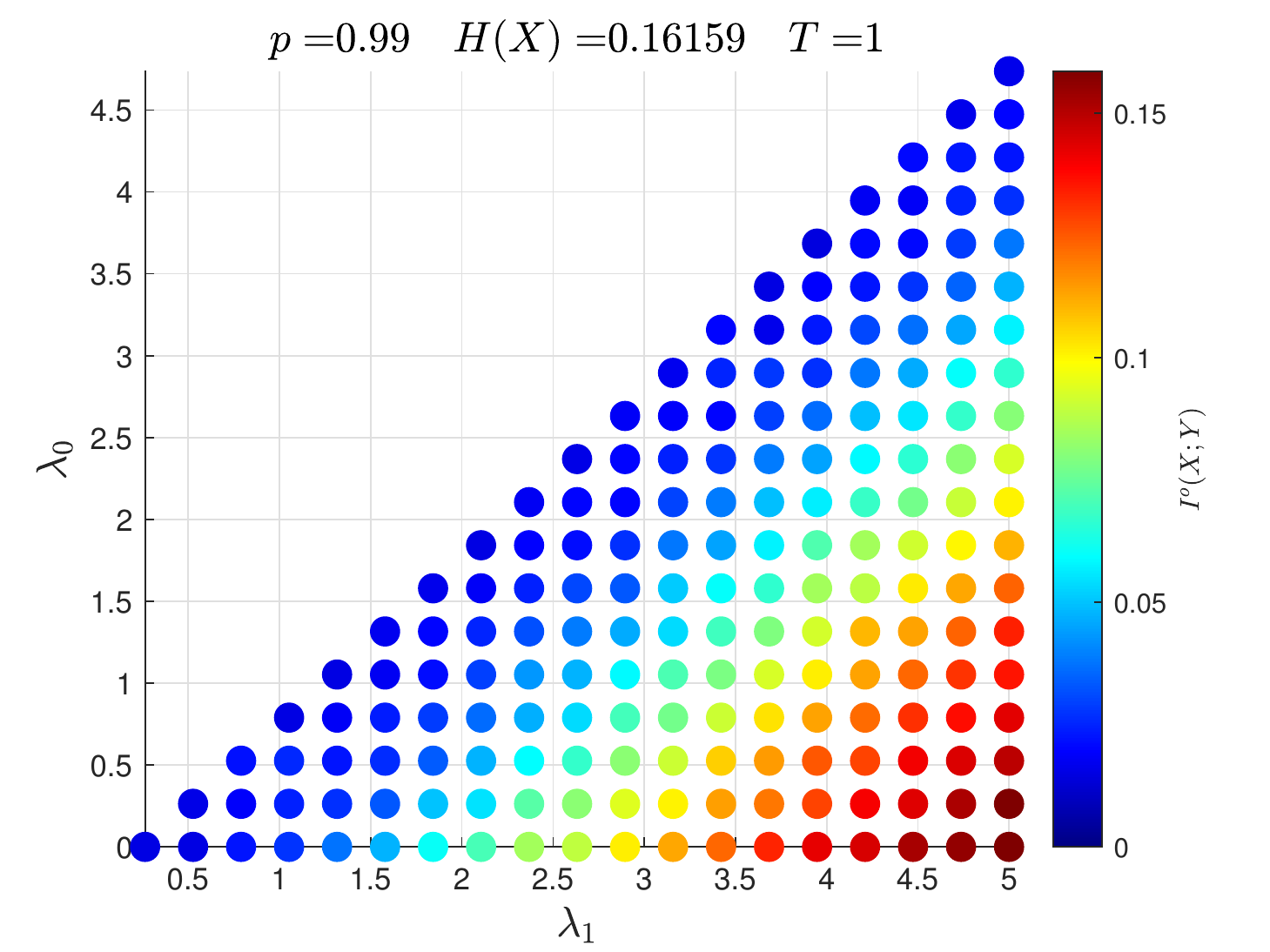}
		\caption{ }
		\label{fig6e}
	\end{subfigure} %
	\begin{subfigure}{.49\textwidth}
		\includegraphics[width=\linewidth]{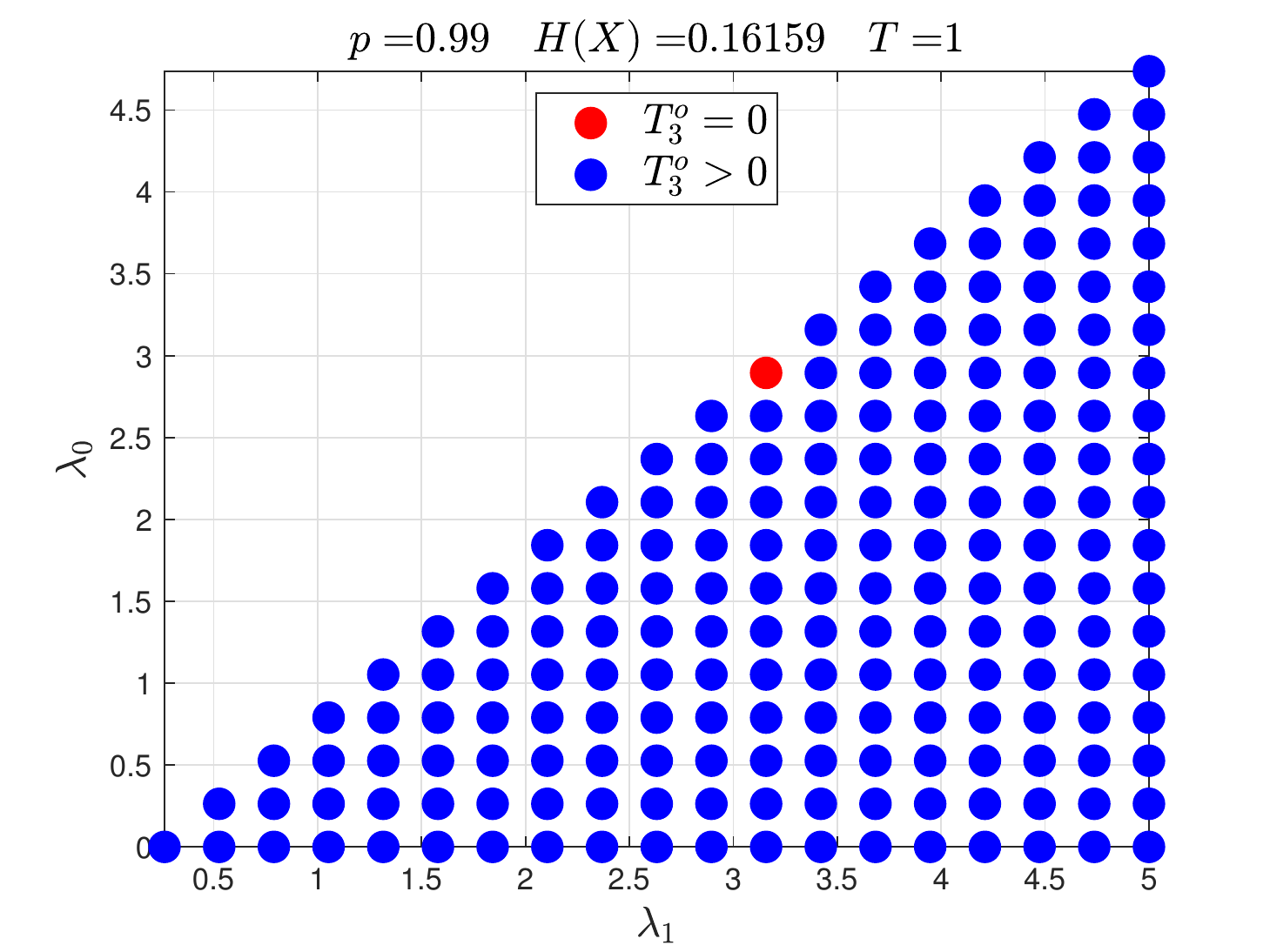}
		\caption{ }
		\label{fig6f}
	\end{subfigure} %
	\caption{Left: $I(X;Y)^O $ vs. $ (\lambda_0 ,\lambda_1 ) $ in the region $\lambda_1 > \lambda_0 $, right: corresponding optimal argument parameter $T_{3}^o$ vs. $ (\lambda_0 ,\lambda_1 ) $ for varying \emph{prior} probabilities $ p $. The search for each optimal argument $T_{3}^o$ for any fixed: $(\lambda_0 , \lambda_1 )$ and $p$ is performed over the line $ (T_1,T_2,T_3) := (\frac{1-T_3 }{2},\frac{1-T_3 }{2},T_3 )$ where $0 \le T_3 \le 1$. \todo[disable,inline]{\texttt{\detokenize{Gauss_Optimal_Alpha3_2.m}}
			\newline   	\texttt{\detokenize{Gauss_MI_2.m}}                     }}
	\label{f6}
\end{figure}

\begin{figure}[ht]
	\begin{subfigure}{.49\textwidth}
		\includegraphics[width=\linewidth]{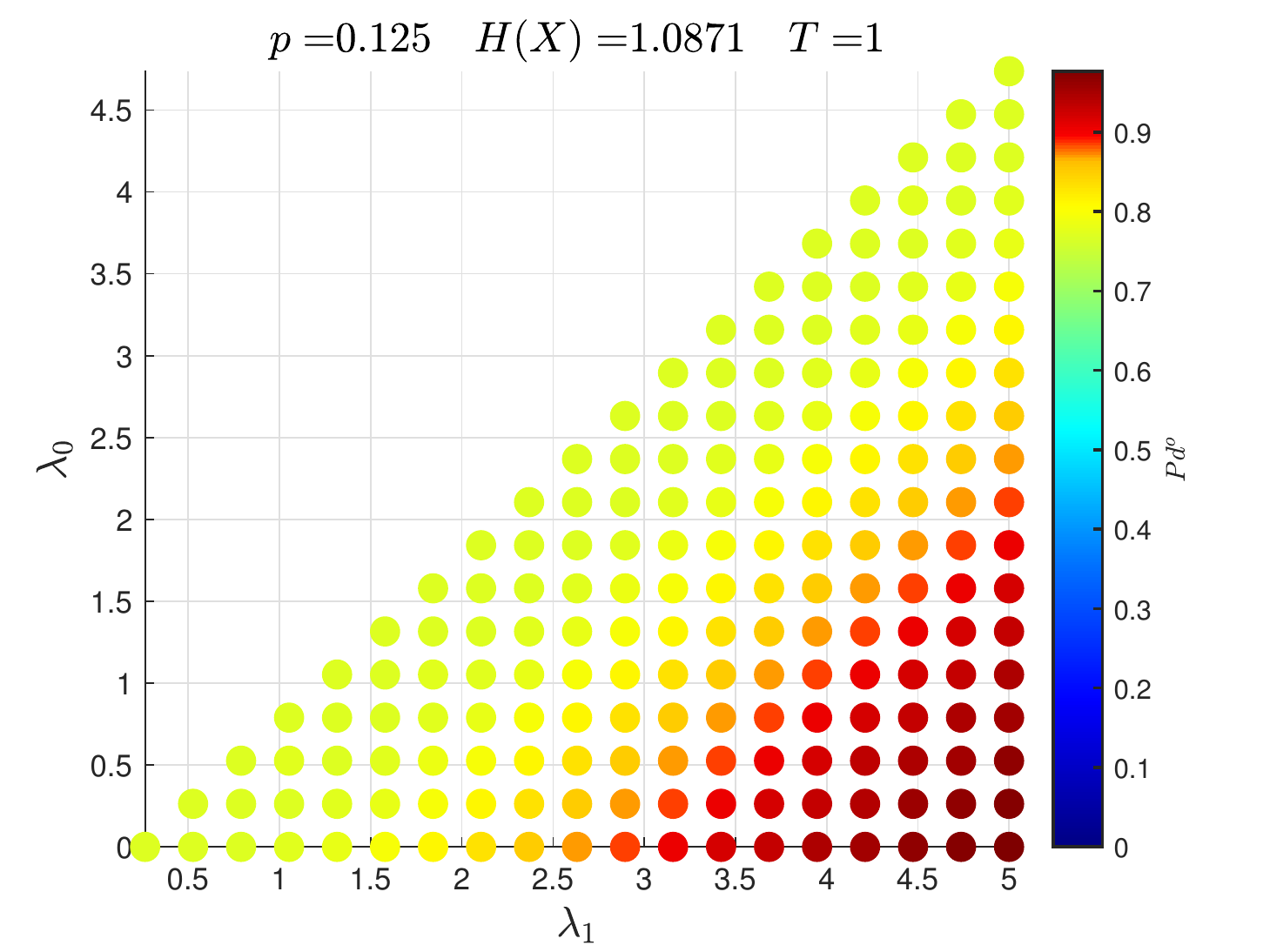}
		\caption{ }
		\label{fig7a}
	\end{subfigure} 
	\begin{subfigure}{.49\textwidth}
		\includegraphics[width=\linewidth]{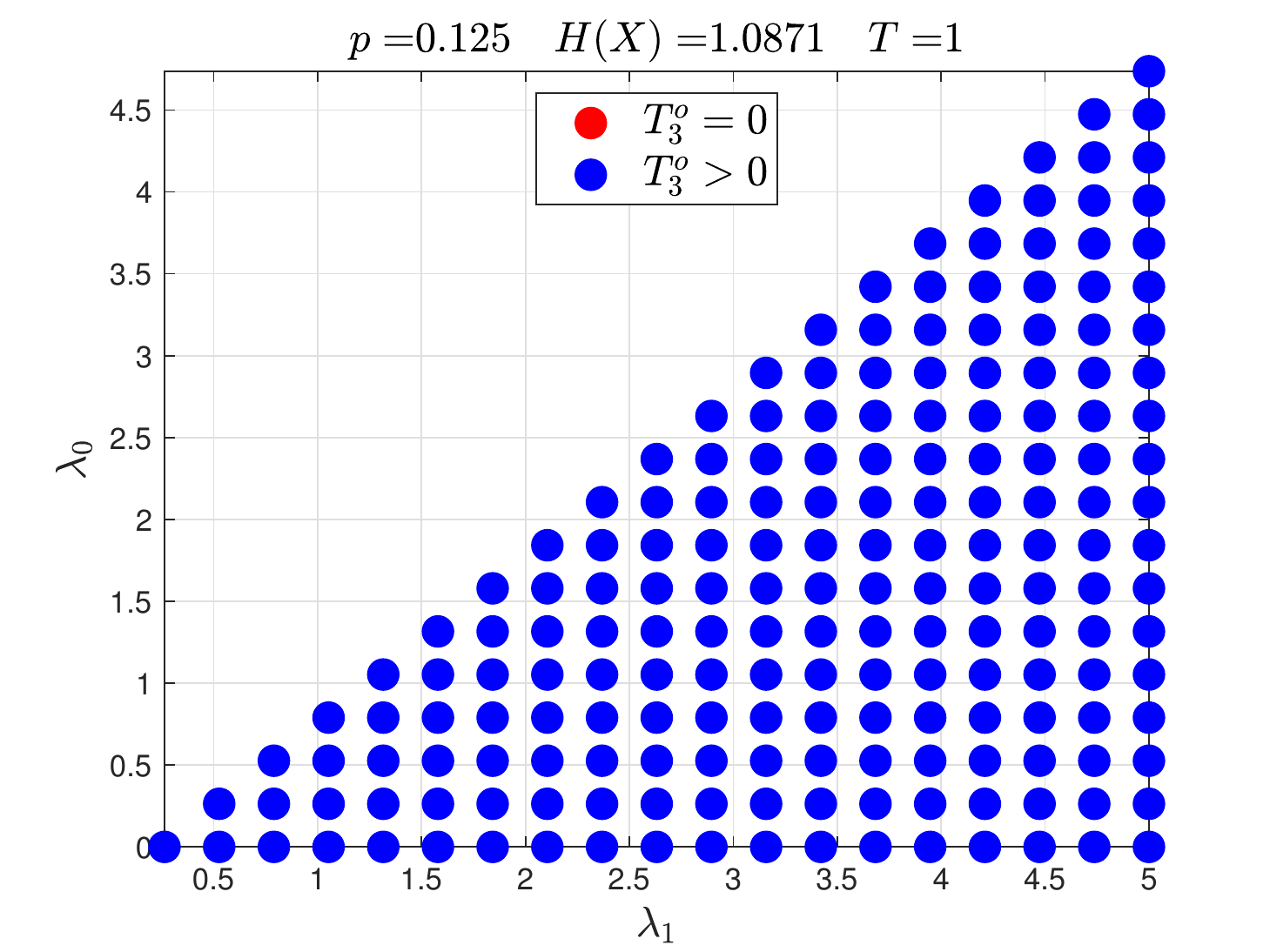}
		\caption{ }
		\label{fig7b}
	\end{subfigure} \\%
	\begin{subfigure}{.49\textwidth}
		\includegraphics[width=\linewidth]{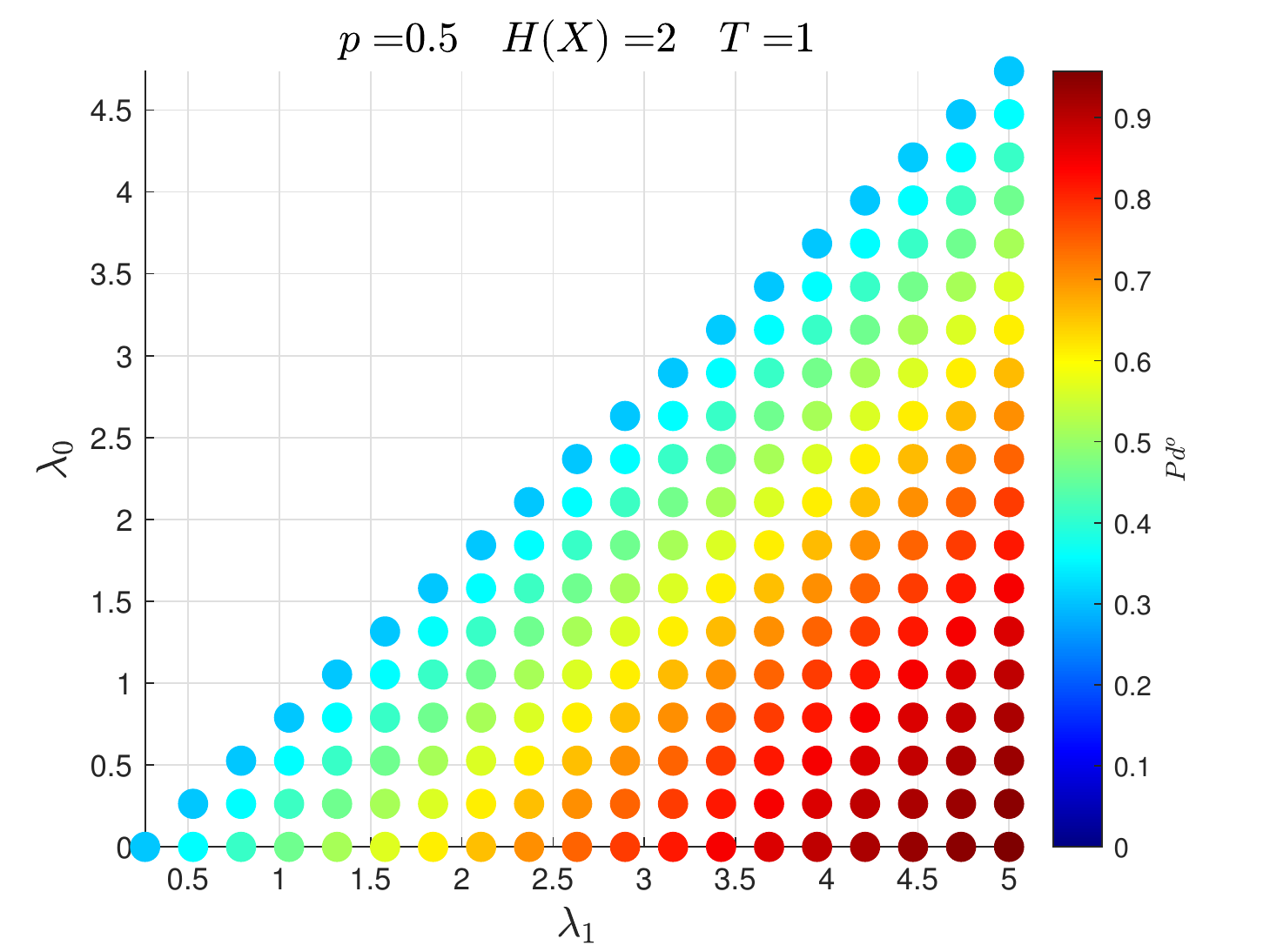}
		\caption{ }
		\label{fig7c}
	\end{subfigure} %
	\begin{subfigure}{.49\textwidth}
		\includegraphics[width=\linewidth]{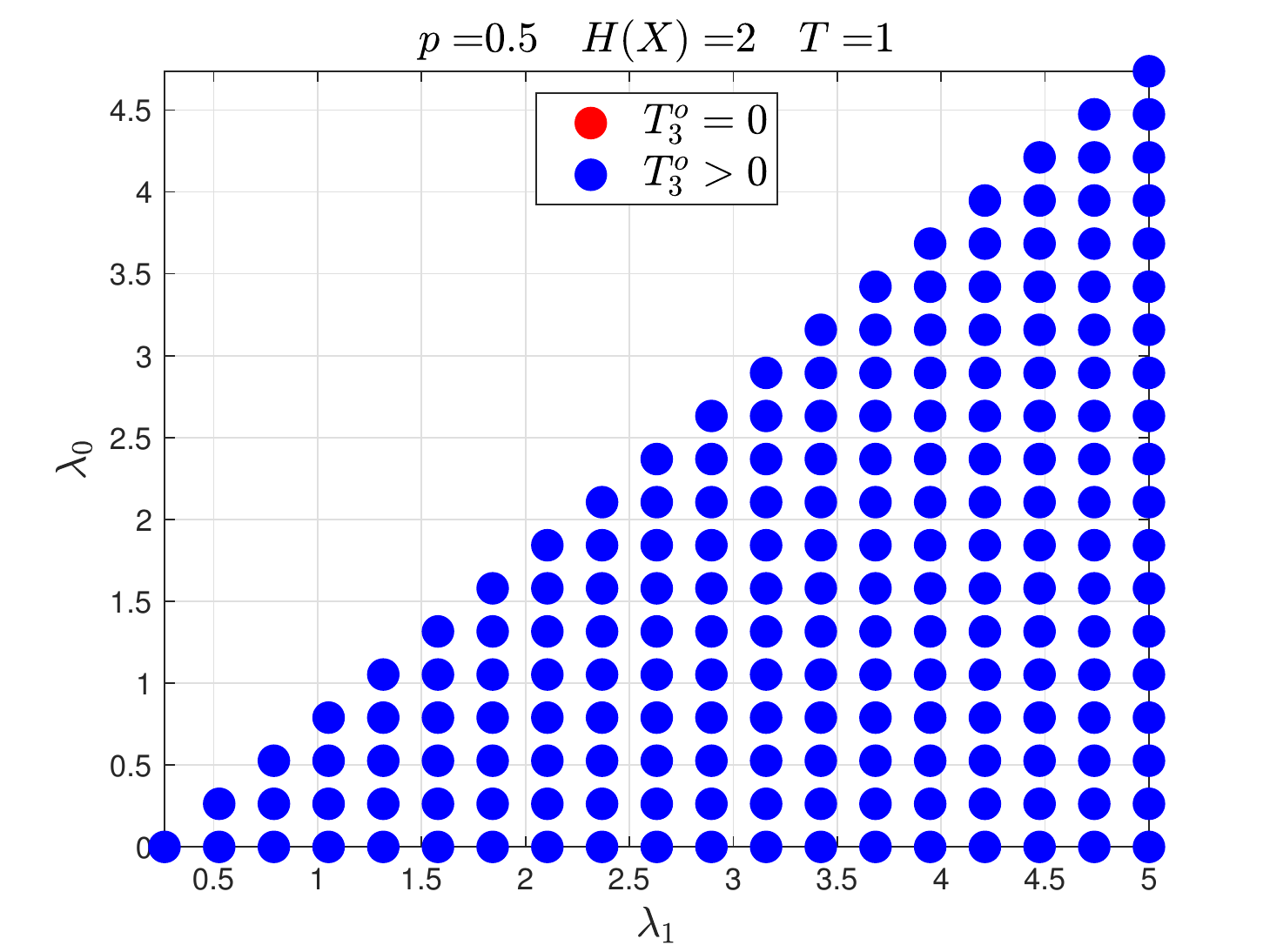}
		\caption{ }
		\label{fig7d}
	\end{subfigure}
	\begin{subfigure}{.49\textwidth}
		\includegraphics[width=\linewidth]{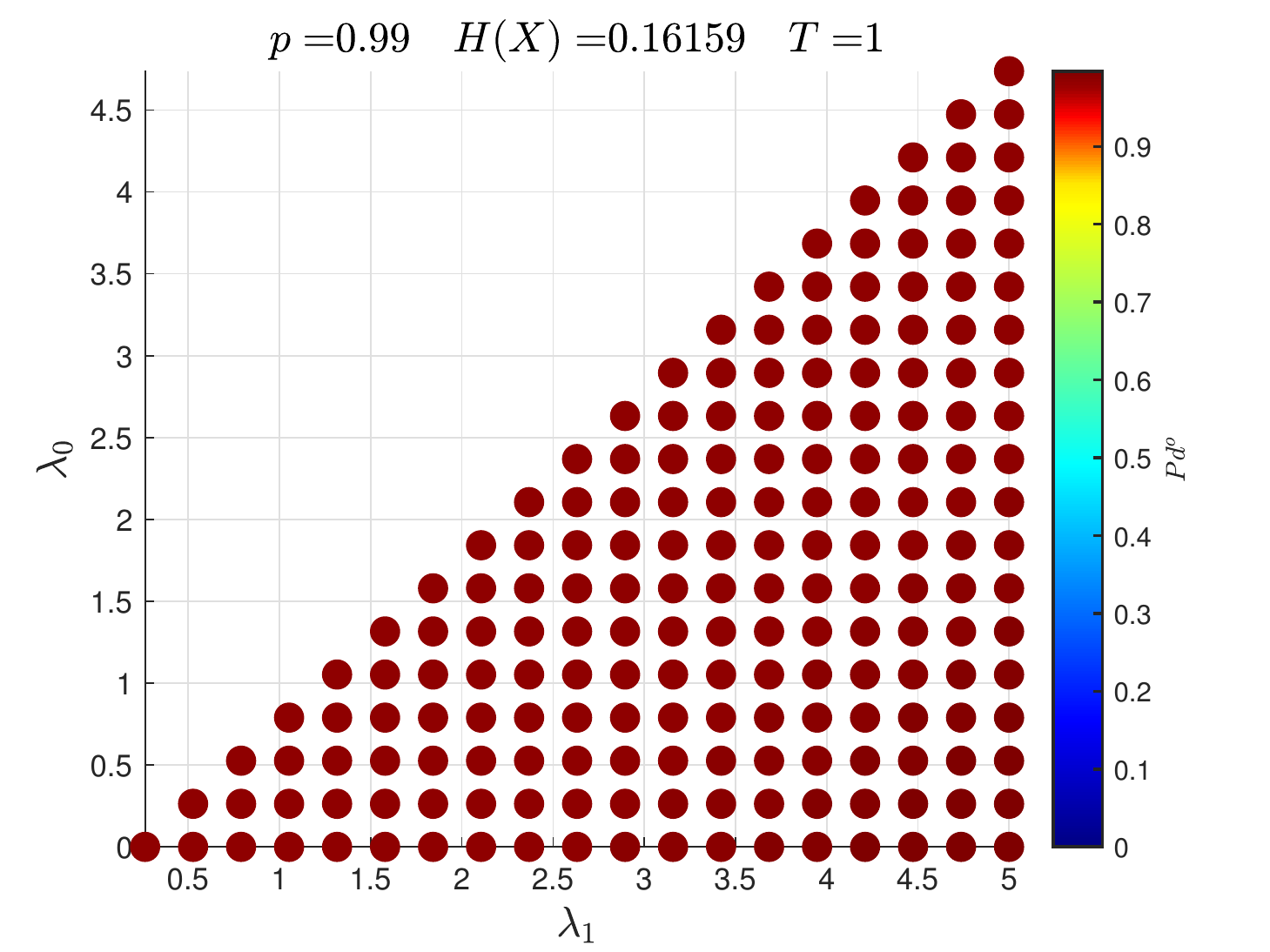}
		\caption{ }
		\label{fig7e}
	\end{subfigure} %
	\begin{subfigure}{.49\textwidth}
		\includegraphics[width=\linewidth]{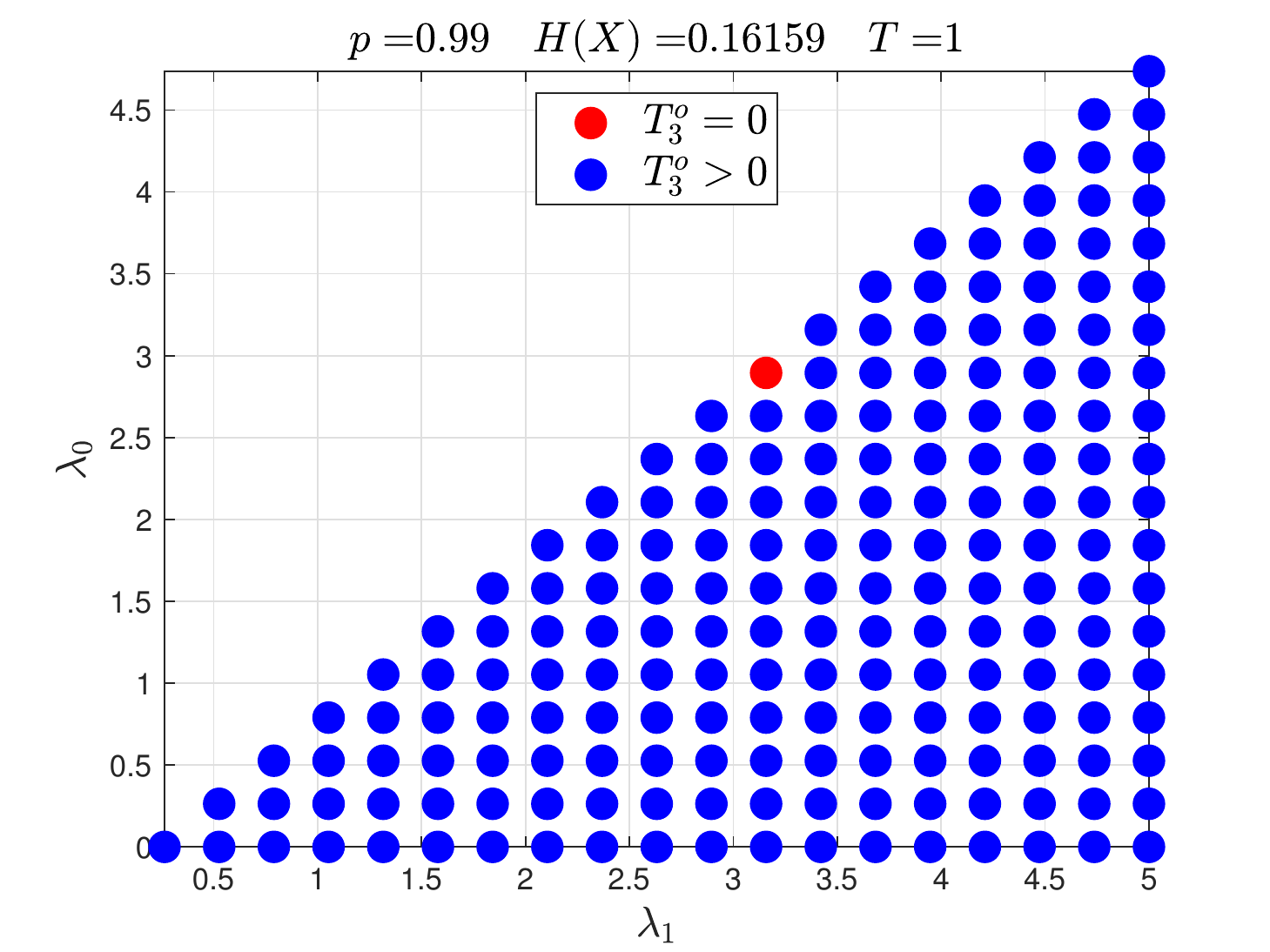}
		\caption{ }
		\label{fig7f}
	\end{subfigure} %
	\caption{Left: $ P_{d}^O $ vs. $ (\lambda_0 ,\lambda_1 ) $ in the region $\lambda_1 > \lambda_0 $, right: corresponding optimal argument parameter $T_{3}^o$ vs. $ (\lambda_0 ,\lambda_1 ) $ for varying \emph{prior} probabilities $ p $. The search for each optimal argument $T_{3}^o$ for any fixed: $(\lambda_0 , \lambda_1 )$ and $p$ is performed over the line $ (T_1,T_2,T_3) := (\frac{1-T_3 }{2},\frac{1-T_3 }{2},T_3 )$ where $0 \le T_3 \le 1$. \todo[disable,inline]{\texttt{\detokenize{Gauss_Optimal_Alpha3_2.m}}
		\newline   	\texttt{\detokenize{Gauss_MI_2.m}}                     }}
\label{f7}
\end{figure}

To further expand our understanding if hybrid sensing remains optimal for a wide range of input parameters $\lambda_0$ and $\lambda_1$ for fixed prior $p$ and time constraint $T=1$, we simulate another Monte Carlo simulation. The input parameter set is $\{(\lambda_0,\lambda_1) \in \mathbb{R}_{+} \times  \mathbb{R}_{+} \big | 0 < \lambda_1 \le 5  \: \mathtt{  and }\:  \lambda_1 > \lambda_0\}$. For each of $(\lambda_0,\lambda_1)$ we compute $400$ values of mutual information by varying $\alpha$ linearly from $0$ to $T=1$ in $400$ steps. For each step $10^5$ samples are used for calculation of differential entropy $H(Y)$. Scatter plots on the left-hand side in fig.(\ref{f6}) illustrates the respective optimal value of $I(X;Y)$ at each input parameter for the prior taking values: $0.125,0.5$ and $0.99$. Whereas the scatter plots on the right-hand side illustrates the corresponding  optimizing argument $(\frac{T-T_3}{2},\frac{T-T_3}{2},T_3)$ where $0 < T_3 < T$. It can be seen that when the two input parameters $\lambda_0$ and $\lambda_1$ are closer (as in the diagonal) the mutual information is near to zero and hybrid sensing is the best sensing strategy; as the two input parameters gets farther (as in the lower right corner in scatter plot) the mutual information gets higher and still the hybrid sensing is optimal. This is true for all three values of the prior. When the same simulation is run for maximizing the Bayes probability of total correct detections $P_d$ the results are shown in scatter plots of fig.(\ref{f7}). $P_d$ is shown on the left scatter plot for each prior. As the input parameters $\lambda_0$ and $\lambda_1$ gets closer (as in the diagonal) the $P_d$ touches the maximum value among the $\{(1-p)^2,  \quad  2p(1-p), \quad p^2\}$. In the lower right corner the $P_d$ is highest as the input parameters are the farthest apart. The right scatter plots illustrates that hybrid sensing is the optimum method from the Bayes detection point of view. It must be noted that even the hybrid sensing is optimal from perspectives of the mutual information and Bayes probability of total correct detection; the optimal arguments from these two metrics are not necessarily appear to be the same. These simulations therefore constitute a counter-example where information theory and detection theory are leading to different optimal solutions. 



  \section{Conclusion} \label{con}
   This work attempts to address the problem of sensor scheduling in a vector Gaussian channel for a two target detection, when a specified structure on scaling matrix is imposed, using criteria of mutual information and Bayesian risk with $0-1$ loss function. From computations, it was found that what is optimal argument under mutual information criterion need not necessarily be optimal under Bayesian inference. It was further found that mutual information is concave in the line  $ (T_1,T_2,T_3):=(\frac{T-\alpha}{2},\frac{T-\alpha}{2},\alpha)$  parametrized by $ 0 \le \alpha \le T $. For any given prior $p$ and given finite time $T$: hybrid sensing is found to be the optimal sensing mechanism for any given time proportions.
   
   There are few open questions that are worth exploring. For example, one can aim to find any counter-example in which mutual information is not Schur concave in the line  $ (T_1,T_2,T_3):=(\alpha \cdot T,(1-\alpha) \cdot T,c)$ parametrized by $ 0 \le \alpha \le 1 $ for some positive fixed constant $c$. One may also consider a problem where both the covariance matrix and the mean of the Gaussian channel are affected by the sensing time proportions and observe which sensing method is suitable for this case.

\bibliographystyle{IEEEtran}
\bibliography{ReferencesGlobal}
\end{document}